\documentclass[final,3p,twocolumn]{elsarticle}
\usepackage{graphicx}
\usepackage{amssymb}
\usepackage{amsmath}
\usepackage{dcolumn}
\usepackage{bm}
\biboptions{numbers,sort&compress}
\journal{Nuclear Physics A}
\begin{document}

\begin{small}
\begin{frontmatter}

\title{Impact of the quenching of shell effects with excitation energy on Nuclear Level Density}
\author[a]{Mamta Aggarwal\corref{cor1}}
\cortext[cor1]{corresponding authors: Mamta Aggarwal, mamta.a4@gmail.com}
\address[a]{Department of Physics, University of Mumbai, Kalina Campus, Mumbai-400098, India}
\begin{abstract}
We investigate the nature and impact of shell effects on nuclear level density (NLD) and particle emission probability as a function of temperature in a microscopic theoretical framework of Statistical Model for nuclei ranging from neutron deficient to neutron rich isotopes of Z$=$ 27$-$35. Critical temperatures are traced for neighbouring even, odd, closed shell and mid$-$shell nuclei which respond to excitations differently due to their varying stability and structural effects. Importance of the shell effects and shell correction energy is reflected significantly in NLD variation which slowly diminishes with increasing excitation energy indicating the quenching of shell effects. The enhancement of LD parameter with the deformation and rotation and the fade out of enhancement with increasing excitations has been shown. The weakening of magicity of N$=$28 near the proton drip line has been observed in the inverse level density parameter 'K'($=$ A/a) and $\beta$ variation though it is usually uncommon to see this effect in excited nuclei. Variation of our calculated NLD for odd $^{69}$As and even $^{70}$Ge  exhibits structural effects and agrees very well with the variation of experimental values. Our evaluated level density parameter 'a' values are compared with RIPL-2 (Reference Input Parameter Library) data which show good agreement. \par
\end{abstract}


\end{frontmatter}

\section{Introduction}
Starting from the pioneering work by Bethe ~\cite{BETHE}, many researchers have been making various efforts to evaluate the nuclear level density (NLD) ~\cite{ERIC,BOHR,IGNATYUK,KATARIA,MTN} and understand it's dependence on key parameters such as excitation energy, angular momentum, particle number, parity, isospin and shell effects. Since NLD is an essential physical input in the statistical model (SM) calculations of the compound nuclear decay, it becomes crucial in determining the nuclear reaction cross sections for various practical applications of astrophysics  ~\cite{WALLA}, nuclear technology, fission or fusion reactor designs. The advancemnent of the experimental methods to extract NLD  ~\cite{KAUSHIK,BALARAM,GOHIL,YKGUP,PROY,PCROUT} and ability to measure the thermodynamic and statistical properties ~\cite{Schiller,Melby,Guttormsen,SCHILLER1} of the excited nuclei have opened up a new horizon to explore the response of the nuclear systems subjected to the thermodynamical changes and consequently the structural changes which impact NLD and the decay modes. \par

The experimental techniques such as the measurement of primary $\gamma$$-$ray spectra ~\cite{SCHILLER1}, direct counting of nuclear levels ~\cite{KATSANOS,Katsanos1,Raman,Mishra}, analysis of neutron resonance spacings ~\cite{Capote}, and SM analysis of particle-evaporation spectra in Heavy-Ion fusion reactions ~\cite{Hasse,Nebbia,Hagel,wada,Gonin,Gonin1,Chbihi,Yoshida,Fabris,Nebbia1,Fineman,Caraley,Charity1,Charity} to compute the nuclear level density provide useful information on various aspects of statistical behaviour of compound nuclei and level densities. Comparison of the spacings of nuclear levels in the neutron resonance data for closed shell and mid-shell nuclei at a similar excitation energy reveal \cite{facchini,BABA,HBABA,TDNEWTON} that the larger shell gap in magic nuclei results in much smaller density of resonances pointing towards a strong dependence of level density on nuclear shell structure ~\cite{manpa,MAK,UKPAL,ACHOUDHARI,SGORIELY}. The nuclear shell effects ~\cite{VM,MASHA,MAPLB}, that arise from one-body interactions, are known to provide extra stability to the closed shell magic nuclei that seem to play a major role in the nuclear phenomena like production of super heavy elements (SHE) ~\cite{ogane,HOFMAN}, fission isomers ~\cite{Polikanov}, superdeformed nuclei~\cite{Janssens} and changing magicity in exotic nuclei ~\cite{YADAV,bastin,GMAIJ17}.  With increasing excitation energy, the damping of the shell effects ~\cite{PCROUT} has significant impact on NLD and other statistical quantites. Knowledge of the critical temperature or excitation energy of a nucleus, at which the nuclear shell effects wash out completely, is important  ~\cite{MPROY,MAPLB,MACO,MAMHIAS,MI,MAJNP,IGNAT,GOOD,ALHA} as it provides valuable insights in the planning of experiments especially in the production of SHE  ~\cite{ogane,HOFMAN} which is one of the most thrust areas of the recent times. However the experimental information on the damping of shell effects is yet scarce although there have been a few efforts in recent times ~\cite{ACHOUDHARI,PCROUT}. \par

Measurements of the damping of shell effects on NLD parameter with excitation energy and the fadeout of collective enhancement of NLD with shape transitions ~\cite{PCROUT,KAUSHPLB,PROY} have indicated the significant influence of shell structure in NLD variations. In a recent theoretical work within a microscopic framework,  we studied \cite{manpa} the impact of spin induced deformation and shape transitions ~\cite{MACO,MAJNP,MA,MAJNP1} on level density \cite{manpa,MAIJMP4,MAIJMP8} where the enhancement of NLD and fluctuation in inverse LD parameter 'K' \cite{manpa} due to shape transition was prominent in mid-shell deformed nuclei but completely absent near the closed shell nuclei \cite{manpa}. This shows that the closed shell and mid-shell nuclei respond to excitations quite differently due to their different shell structure and stability. Hence one must consider the shell effects while evaluating NLD for which one needs an accurate value of shell correction energy $\delta$E$_{shell}$ ~\cite{VM} which is usually several MeVs in doubly magic nuclei and impacts NLD and other the statistical quantities significantly. Effect of shell correction on the LD parameter over a range of excitation energy where the effect of damping is significant, has been measured recently ~\cite{PCROUT}. Their experimental results show that the shell correction is indeed necessary to explain the data which is much more pronounced in the closed shell region. \par
In this work, we use Microscopic Statistical Model (MSM)~\cite{MACO,MAPLB} and Macroscopic-Microscopic approach using the triaxially deformed Nilsson-Strutinsky model (NSM) \cite{MASHA} to study shell effects on NLD. In this model ~\cite{MA}, the excited compound nuclei are described as the thermodynamical system of fermions incorporating the deformation and temperature and rotational degrees of freedom. The deformation effects increase the number of particles in the classically forbidden region below the continuum threshold  which contribute to the pairing interactions. The scattering of valence particles near the Fermi level due to pairing correlations influence the shell structure and the decay modes. The imapct of pairing effects and the damping of pairing effects with the excitation energy ~\cite{MTN} on the level density parameter is noteworty as has been shown by Rajasekaran et al. \cite{MTN}.  However, the effects of the pairing become unimportant especially for temperatures above T$\approx$ 0.8 as seen in Ref. ~\cite{MTN}. Since the temperatures considered in the present work are around 1 MeV - 3 MeV where the pairing correlations are negligible, our present formalism works well and hence the pairing corrections have not been included in the calculations. Here we focus on shell efects, deformation and temperature degrees of freedom. Since the intrinsic states in a deformed nucleus give rise to rotational bands which enhance the level density over that of a spherical nucleus, we have also included the rotatioal degree of freedom in few calculations. A systematic study on the shell effects and thier impact on NLD is presented. Our results agree well with the RIPL data and the other available experimental data.  \par

\section{Theoretical Description}
The neutron emission spectra (P) for a deexciting compound nuclear system can be obtained by the number of neutrons emitted within an energy interval E$_n$ and E$_n$ +dE$_n$ using the following equation ~\cite{BLAT}

\begin{equation}
dN(E_n) = C E_n \rho (U_{th}) dE_n,
\label{yld}
\end{equation} 
where $\rho$(U$_{th}$) is the nuclear level density of the residual nuclear system. The residual thermal energy U$_{th}$ of the daughter nucleus is found by 
\begin{equation}
U_{th} = E^* - S_n - E_{rot} - E_n
 \label{ures}
\end{equation} 
where E$^*$ is the total excitation energy available to the system due to the reaction process which is shared between the various degrees of freedom like the outgoing neutron energy E$_n$, neutron separation energy S$_n$ and the rotational energy E$_{rot}$.  C is the normalisation constant. The average kinetic energy of the outgoing neutron E$_n$ which dependes upon the energy available to the deexciting nucleus, is varied from 0 to 8 MeV for a given spin and temperature. According to Eqs. \ref{yld} and \ref{ures}, When U$_{th}$ is large, neutron emission probability is large. Obviously U$_{th}$ is large when E$_{rot}$ and S$_n$ are small at a given T. In fact, for neutron emission to occur,  E$^*$ should be greater than (E$_{rot}$ +  S$_n$). \par
The nuclear level density ($\rho$(U$_{th}$)) at excitation energy (U$_{th}$) of the residual nucleus is obtained using the expression given below : \par
\begin{equation}
\rho(U_{th}) = {(\hbar^2/2 \Theta)^{3/2} (2I +1) {\sqrt a} exp(2 \sqrt{aU_{th}}) \over 12(U_{th}+T)^2},
\label{nldr}
\end{equation}
where $\Theta$ is the moment of inertia and 'a' is the single particle level-density parameter calculated as a function of M and T
\begin{equation}
 a(M,T) = S^2 (M,T) / 4 U_{th}(M,T)   
\end{equation}
The moment of inertia is obtained using the expression ~\cite{BLAT,HM}
\begin{equation}
\Theta_1 = \hbar^2 I {(dE_{rot}  /dI)}^{-1}
\label{theta}
\end{equation}
\begin{equation}
\Theta_2 = \hbar^2 {(d^2E_{rot} /dI^2)}^{-1}
\label{thetay}
\end{equation}
Eq.\ref{thetay} is used only when there is band crossing. For a non-rotating system, nuclear level density  is given as
\begin{equation}
\rho(U_{th}) = {\sqrt\pi exp(2\sqrt{aU_{th}}) \over 12 U_{th}(aU_{th})^{-1/4}} ,
\label{nldnr}
\end{equation}
where U$_{th}$ = E$^*$ - S$_n$ - E$_n$. \\
E$^*$ and E$_{rot}$ are computed using our formalism MSM ~\cite{MACO,MAPLB}. The ground state separation energy (S$_N$$^{cor}$) is calculated by using the Macroscopic -Microscopic approach using the NSM ~\cite{MASHA,MAPLB} where we include the shell effects arising due to the nonuniform distribution of nucleons through the  Strutinsky's shell correction $\delta$E$_{shell}$ ~\cite{VM,MASHA} along with the deformation energy E$_{def}$ (obtained from the surface and coloumb effects) to the macroscopic binding energy BE$_{LDM}$ obtained from the LDM mass formula ~\cite{PM}     
\begin{eqnarray}
S_N^{cor}(Z,N,\beta,\gamma)= BE_{gs}(Z,N,\beta,\gamma)-\nonumber\\
BE_{gs}((Z,N-1,\beta,\gamma)
\label{sncor}
\end{eqnarray}
where 
\begin{eqnarray} 
BE_{gs}(Z,N,\beta,\gamma) = BE_{LDM}(Z,N)-\nonumber\\
E_{def}(Z,N,\beta,\gamma)-\delta E_{shell}(Z,N,\beta,\gamma)
\end{eqnarray}
where $\delta$E$_{shell}$ = - $\delta$BE$_{shell}$ is the shell correction to energy which is added to the macroscopic energy of the spherical drop BE$_{LDM}$ ~\cite{PM} along with the deformation energy E$_{def}$ gives the total energy  BE$_{gs}$ ~\cite{MASHA,MAPLB} corrected for microscopic effects of the nuclear system. It is computed as ~\cite{MASHA}
\begin{equation}
\delta E=\sum_{i=1} ^A \epsilon_i- \tilde E
\end{equation}
where the first term is the shell model energy in the ground state and the second term is the smoothed energy. 
The macroscopic separation energy (S$_N$$_{LDM}$) which does not incorporate shell effects is obtained as 
\begin{equation}
S_N(LDM)= BE_{LDM}(Z,N)- BE_{LDM}(Z,N-1)
\label{snldm}
\end{equation}
The excitations in a nucleus are included through the grand canonical partition function Q($\alpha$$_Z$, $\alpha$$_N$, $\beta$, $\gamma$) for a hot rotating nucleus with N neutrons and Z protons ~\cite{MNV,MORET} at a temperature T=1/$\beta$.
\begin{equation}
Q(\alpha_Z,\alpha_N,\beta',\gamma')=\sum exp(-\beta'E_i+\alpha_Z Z_i+\alpha_NN_i+\gamma' M_i)
\end{equation}
where the Lagrangian multipliers $\alpha$$_Z$, $\beta$', $\gamma$' conserve the particle number $Z$, $N$, total energy $E$ and angular momentum $M$ of the system and are fixed by the saddle point equations (see ~\cite{MAPLB} for the detailed formalism). The conservation equations in terms of single particle levels for the protons $\epsilon$$_i$$^Z$ with spin projection m$_i$$^Z$ and for neutrons  $\epsilon$$_i$$^N$ with spin projection  m$_i$$^N$ are,
\begin{equation}
 < Z >  = \sum n_i^Z, \;\;\;\;\;\; < N >  = \sum n_i^N  
\end{equation}
\begin{equation}
 <E(M,T)>  = \sum n_i^Z \epsilon_i^Z + \sum n_i^N \epsilon_i^N 
\end{equation}
\begin{equation}
 < M >  = \sum n_i^Z m_i^Z + \sum n_i^N m_i^N 
\end{equation}
where n$_i$ is the occupation probability 
\begin{equation}
 \sum n_i = \sum [1 + exp(-\alpha +\beta' \epsilon_i - \gamma' m_i)]^{-1},
\end{equation}
The single particle level schemes are obtained by the diagonalisation of the triaxially deformed Nilsson Hamiltonian in the cylindrical basis states ~\cite{EIS,GS} with the Hill-Wheeler ~\cite{HW} deformation parameters ($\beta$,$\gamma$). The excitation energy E$^*$(= E(M,T) - E(0,0)) and rotational energy E$_{rot}$ of this system are computed as 
\begin{equation}
E^* = \sum n_i^Z \epsilon_i^Z + \sum n_i^N \epsilon_i^N  - \sum \epsilon_i^Z - \sum\epsilon_i^N.
\label{exener}
\end{equation}
\begin{equation}
E_{rot}(M) = E(M,T) - E(0,T).
\end{equation}
As T$\rightarrow$ 0, E$_{rot}$ corresponds to the yrast energy.\par
The entropy (S) of the system is computed using 
\begin{equation}
S_Z = -\sum [(n_i^Z ln n_i^Z +(1- n_i^Z)ln(1- n_i^Z)]
\end{equation}
\begin{equation}
S_N = -\sum [n_i^N ln n_i^N +(1- n_i^N) ln( 1- n_i^N)]   
\end{equation}
To evaluate the equillibrium deformation and shape of the excited nuclear systems for a given Z, N and T, we minimize the appropriate free energy $F$ $=$ $E$ $-$$TS$ with respect to  Nilsson deformation parameters $\beta$ and $\gamma$ using the equation    
\begin{eqnarray}
F(Z,N,\beta,\gamma,T) = E(Z,N,\beta,\gamma,M,T) - \nonumber\\
T * S(Z,N,\beta,\gamma,M,T)
\end{eqnarray}
where the total energy E of the excited nuclear system at a given T and M is obtained as  
\begin{equation}
E(Z,N,\beta,\gamma, M,T)= - BE_{gs}(Z,N) + E^*(M,T,\beta,\gamma)
\end{equation} 
where we blend the excitation energy E$^*$ obtained from the statistical Model (MSM) ~\cite{MAPLB} to the ground state energy BE$_{gs}$ calculated using macroscopic-microscopic approach~\cite{MA,MAPLB} that includes the shell correction~\cite{VM} and deformation effects. We trace F minima with respect to intrinsic  deformation parameters $\beta$ and $\gamma$ that also describe the orientation of the nucleus with respect to its rotation axis. $\beta$ and $\gamma$ corresponding to F minima are used for all the calculations. The angular deformation parameter $\gamma$ ranges from -180$^o$ (oblate with symmetry axis parallel to the rotation axis) to -120$^o$ (prolate with symmetry  axis perpendicular to rotation axis) and the axial deformation parameter $\beta$ ranges from 0 to 0.4 in the steps of 0.01. Our formalism has been quite successful in predicting the level density parameter and level density quite accurately ~\cite{MAK,manpa,MPROY} along with the structural transitions in ground and excited states ~\cite{MAPLB,MACO}. \par 

We choose closed shell nucleus Z$=$28 and odd $-$ Z nuclei (Z$=$ 27-35) with N ($=$28$-$52) ranging from neutron deficient to neutron rich isotopes for the present study. The choice of well deformed nuclei $^{60-84}$As, $^{62-87}$Br and doubly magic nuclei $^{56,78}$Ni have helped to demonstrate the shell effects in NLD variation ~\cite{ACHOUDHARI,PCROUT} which, in this work, are found quite prevalent even at quite high excitation energies. Rotational degree of freedom is included by varying angular momentum from 0 to 40 $\hbar$. \par

\section{Results and Discussion}

\begin{figure}[htb]
\centering
\includegraphics[width=0.42\textwidth]{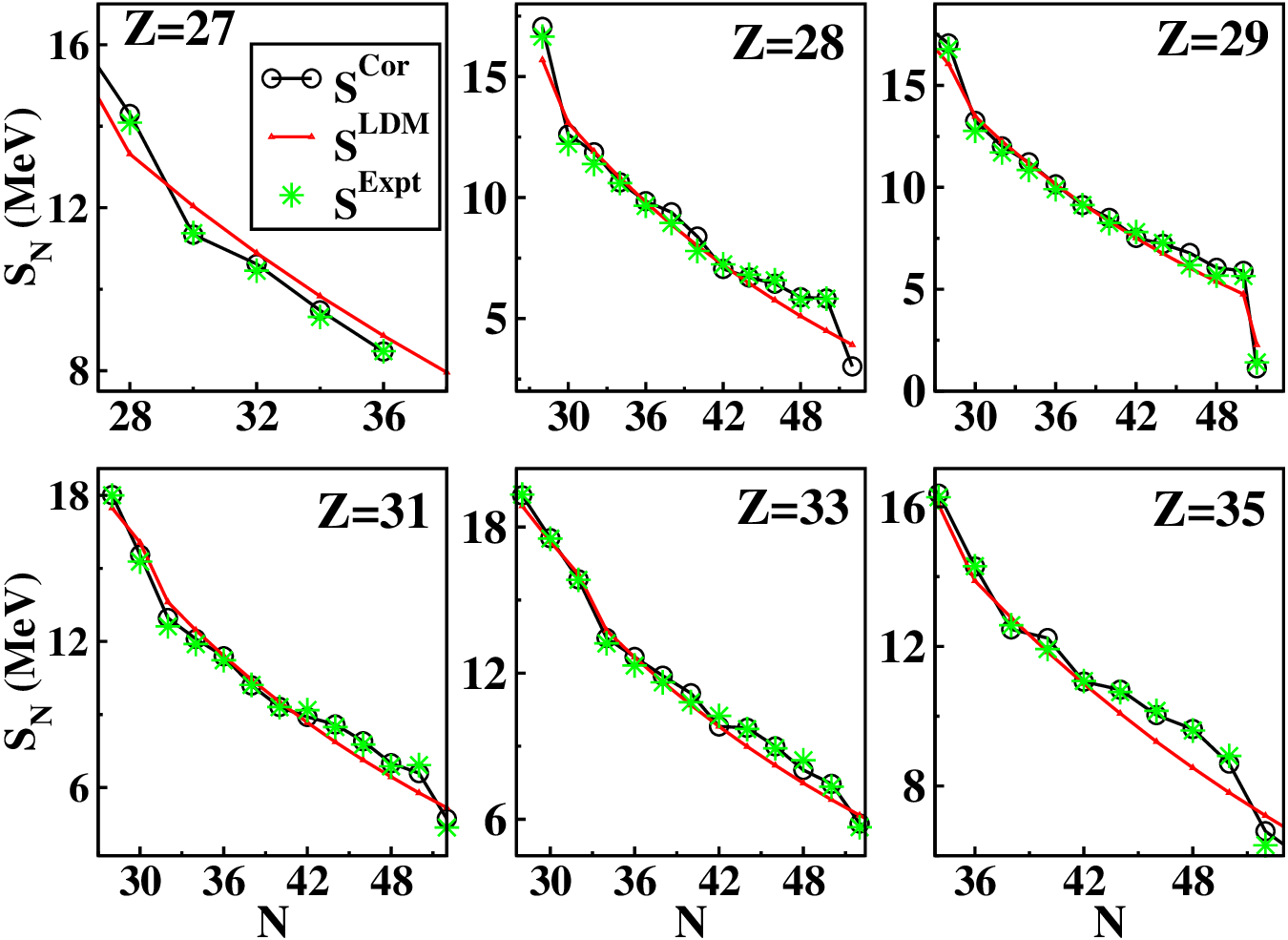}
\caption{(Colour online) Variation of neutron separation energy  S$_N$$^{cor}$, S$_N$$^{LDM}$ and the available data of S$_{N}$ vs. N}
\label{gsn}
\end{figure}
\begin{figure}[htb]
\centering
\includegraphics[width=0.42\textwidth]{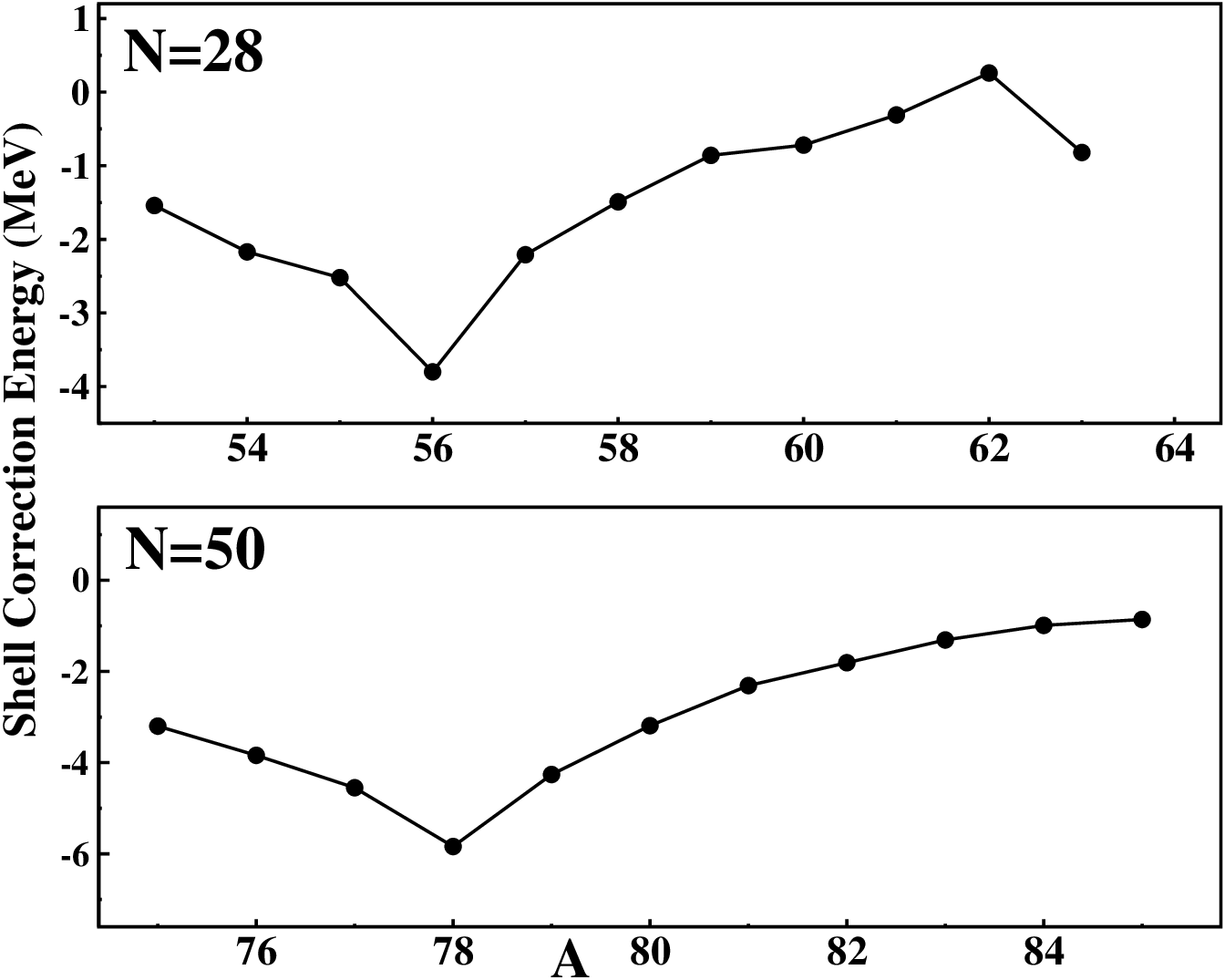}
\caption{(Colour online) Shell correction  $\delta$E$_{shell}$  for (a) N$=$28 and (b) N $=$ 50 isotones.}
\label{gshecor}
\end{figure}
To study the level density and particle evaporation spectra, the total excitation energy E$^*$ available to the compound nuclei As$^*$, Br$^*$, Ni$^*$ due to reaction processes has been evaluated for temperatures T $=$ 1.0 to 3.0 MeV. Since the residual excitation energy (U$_{th}$) after the compound nucleus deexcitation by particle evaporation has significant dependence on the separation energy S$_N$, the accurate determination of the value of S$_N$ is essential. Although one may  use the experimental data of separation energy in such calculations but the experimental data for various nuclear systems especially near the drip lines is still scarce and one has to rely on the theoretical predictions. Our formalism ~\cite{MASHA,MAIJMP4,MA} that has been able to successfully reproduce the measured values of separation energy over a wide range of nuclei, has been used here. Our calculated neutron separation energies  S$_N$$^{cor}$ and  S$_N$$^{LDM}$, computed using the Eqs. \ref{sncor} and \ref{snldm}, are plotted in Fig. \ref{gsn} for Z$=$28 and odd-Z nuclei with Z$=$ 27 to 35 along with the experimental values for comparison. Excellent agreement of data with  S$_N$$^{cor}$ exhibits the accuracy of our calculations. The impact of shell correction is evident in the peaks of S$_N$$^{cor}$ at magic numbers and sudden drop in  S$_N$$^{cor}$ just after the magic neutron numbers N$=$28 and 50 which is not seen in S$_N$$^{LDM}$ which varies smoothly with particle number. Fig. \ref{gshecor} shows the shell correction energy for N$=$ 28 and 50 isotones which is $\approx$ 4 $-$ 6 MeV for doubly magic nuclei in this region and is large enough to impact NLD in a major way. \par
\begin{figure}[htb]
\centering
\includegraphics[width=0.42\textwidth]{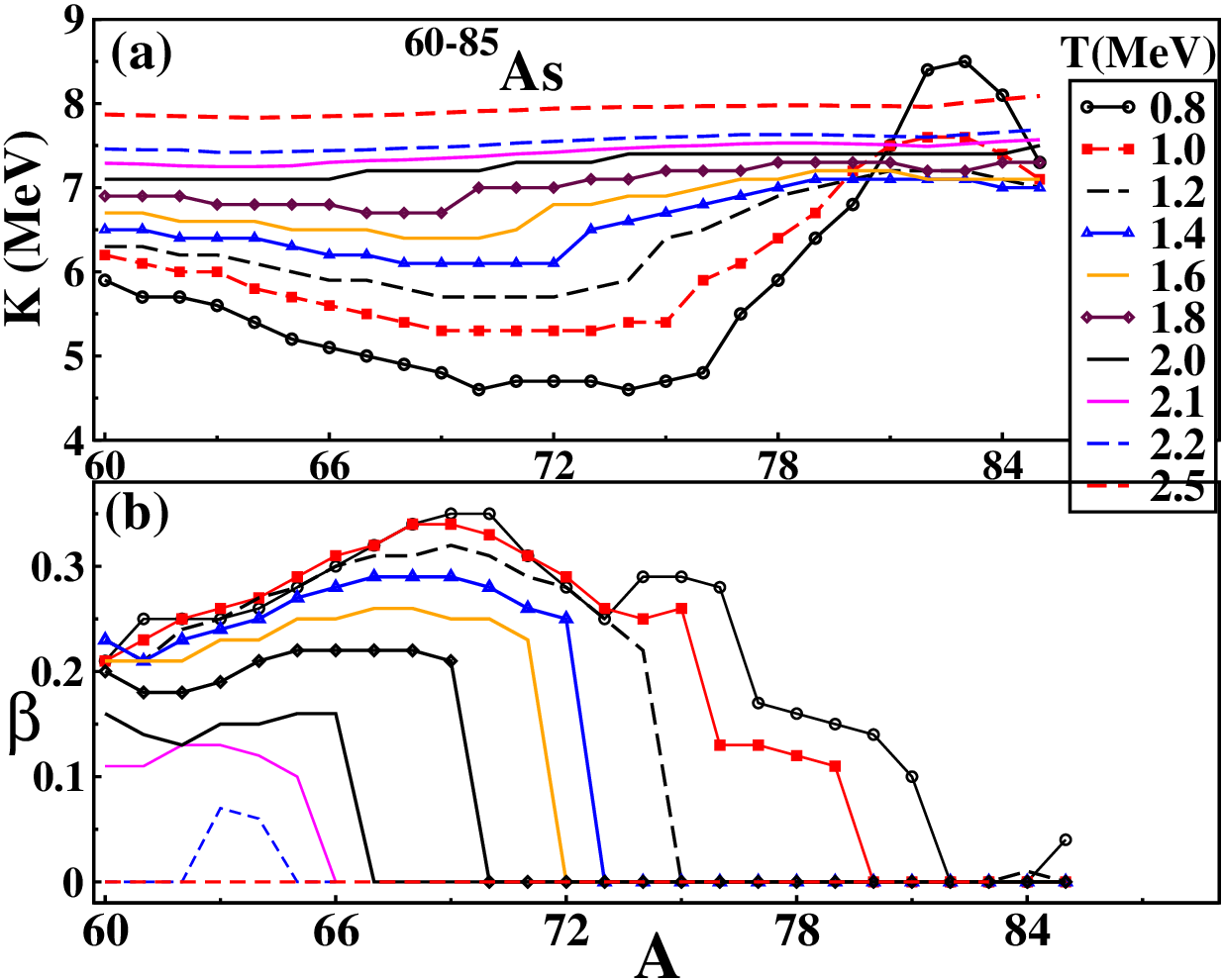}
\caption{(a) Inverse Level density parameter 'K=A/a'and (b) deformation ($\beta$) vs. A for As isotopes $^{60-84}$As at different temperatures.}\label{kvsAAs}
\end{figure}
\begin{figure}[htb]
\centering
\includegraphics[width=0.42\textwidth]{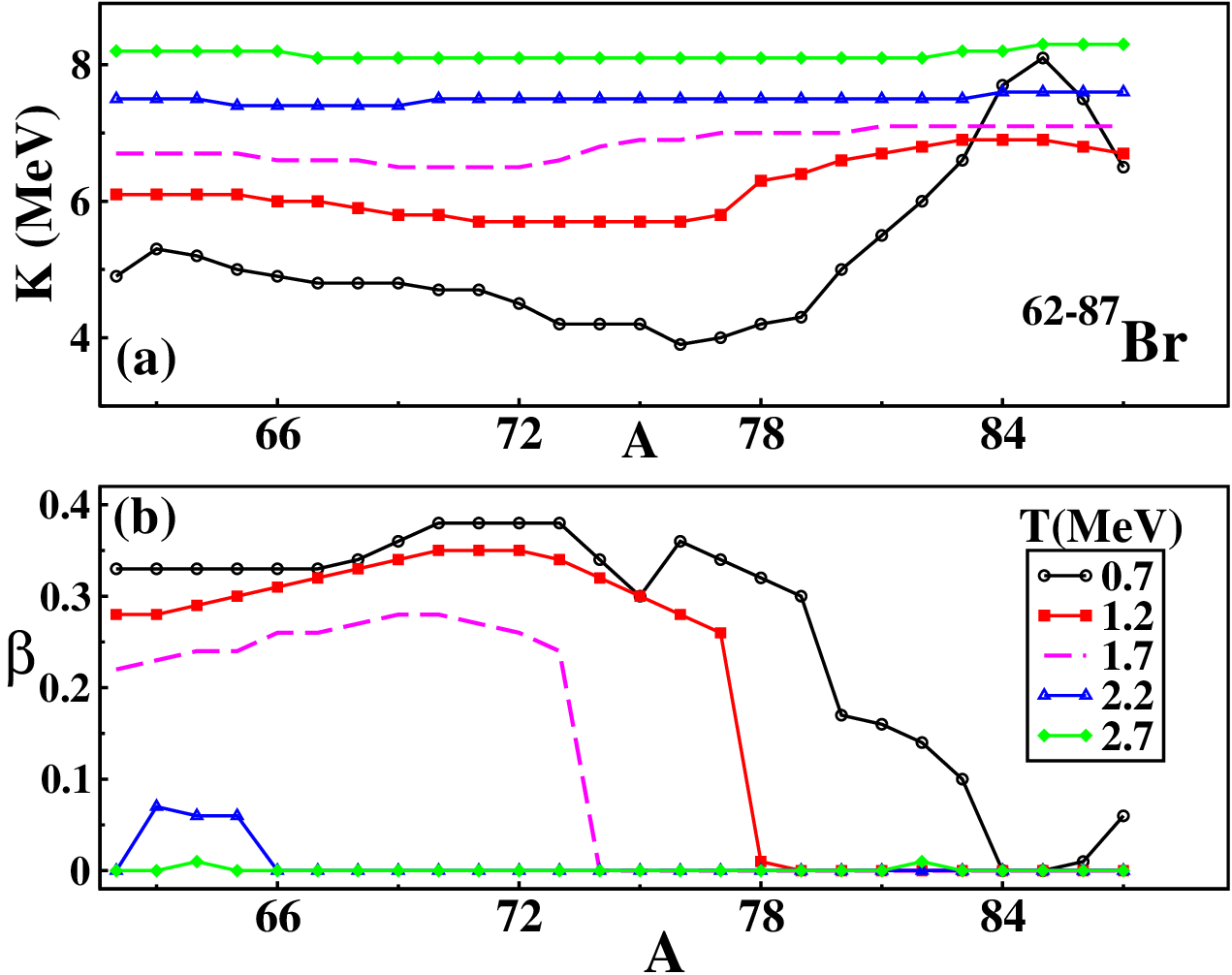}
\caption{(a) Inverse Level density parameter 'K' and (b) deformation ($\beta$) vs. A for Br isotopes $^{62-87}$Br at different temperatures.}\label{kvsABr}
\end{figure}
Inverse level density parameter 'K' for nuclei ranging from the neutron deficient to neutron rich isotopes of $^{60-84}$As (Z$=$33) and $^{62-87}$Br (Z$=$35) are plotted in Fig. \ref{kvsAAs} and \ref{kvsABr}. At low temperatures, the fluctuations in 'K' with prominent peak at shell closures are indicative of the shell effects that start diminishing with increasing T and get completely smoothed out at higher temperatures indicating the quenching of shell effects with excitation. The critical temperatures are different for different isotopes, where the equillibrium deformation drops to zero and the nuclei attain sphericity. At critical tempertaure, the LD parameter 'a'($=$A/K) varies smoothly with A and approaches almost a constant value around A/8 showing the complete wash out of shell effects.\par
\begin{figure}[htb]
\centering
\includegraphics[width=0.42\textwidth]{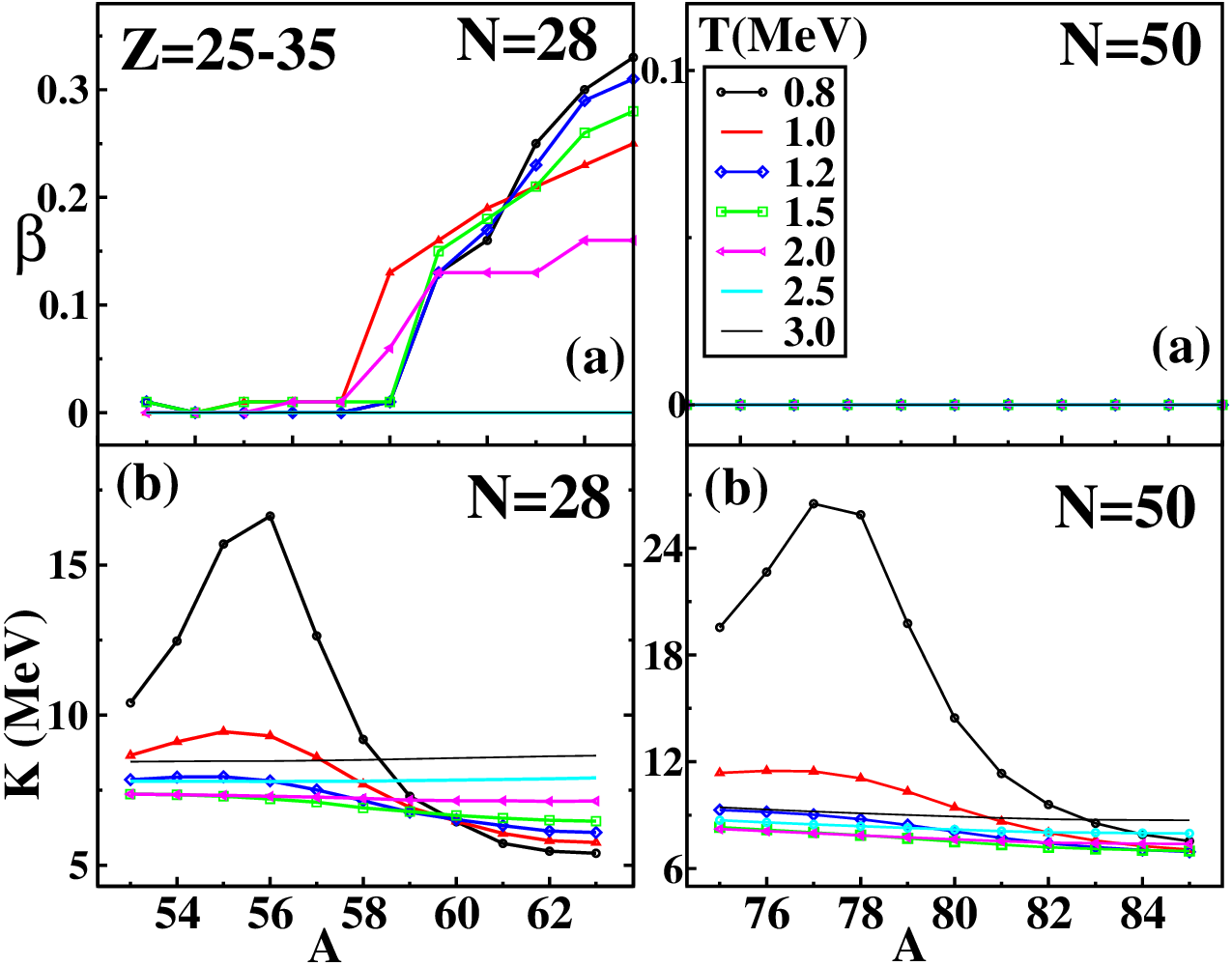}
\caption{(a) $\beta$ (b) Inverse Level density parameter 'K' for N=28 and 50 isotones of Z= 25 to 35 at T=0.8 to 3.0 MeV.}
\label{kdef2850}
\end{figure}
In Fig. \ref{kdef2850}, we have shown the plots of K and $\beta$ vs A for N$=$28 and 50 isotones of Z$=$25-35. The fluctuations in K at low temperature, indicating shell effects, get smoothed out with increasing excitation energy above the critical temperature and indicate the damping of shell effects.  At low temperature, K fluctuates and increases with A while moving towards a shell closure and decreases with A while going away from the shell closure. After the shell effects get quenched at around critical temperatures, the inverse level density parameter K increases smoothly with temperature  \cite{BIJY} as also seen in our work  ~\cite{MPROY} where our calculated values showed reasonable agreement with the average variation of K observed in the experimental data ~\cite{MPROY}. \par
A closer inspection of Fig. (\ref{kvsAAs}, \ref{kvsABr})(b) and \ref{kdef2850} (a) showing $\beta$ vs. A, reveals that the critical temperature (T$_c$) is higher (T$_c$ $=$ 2$-$2.3 MeV) on the neutron deficient side of N=28 isotones than on the neutron rich side (T$_c$ $=$1.0$-$1.6 MeV). Also the N=50 isotones (Fig. \ref{kdef2850}) are all spherical with $\beta$ $=$0 for all the temperatures whereas the neutron deficient isotones of N=28 are well deformed with $\beta$ $\approx$ 0.2 for T upto 2 MeV. Since  N$=$28 is a magic number, large deformation in N$=$28 isotones points towards the weakening of magicity ~\cite{GMAIJ17} of N$=$28 in neutron deficient region. Disappearance of conventional magic numbers and the emergence of new magic numbers near the drip lines is an emerging area of interest  ~\cite{GMAIJ17,Iwasaki,Watanabe}. However, the collapse of magiciy is mostly discussed in ground state nuclei and it is quite uncommon to discuss it in excited nuclei as done here. It is interesting to see the uncommonly seen weakning of magicity of N$=$28 visible in excited nuclei lying towards the neutron deficient side (A$=$ 58 to 63). \par
\vspace{0.6cm}
\begin{figure}[htb]
\centering
\includegraphics[width=0.46\textwidth]{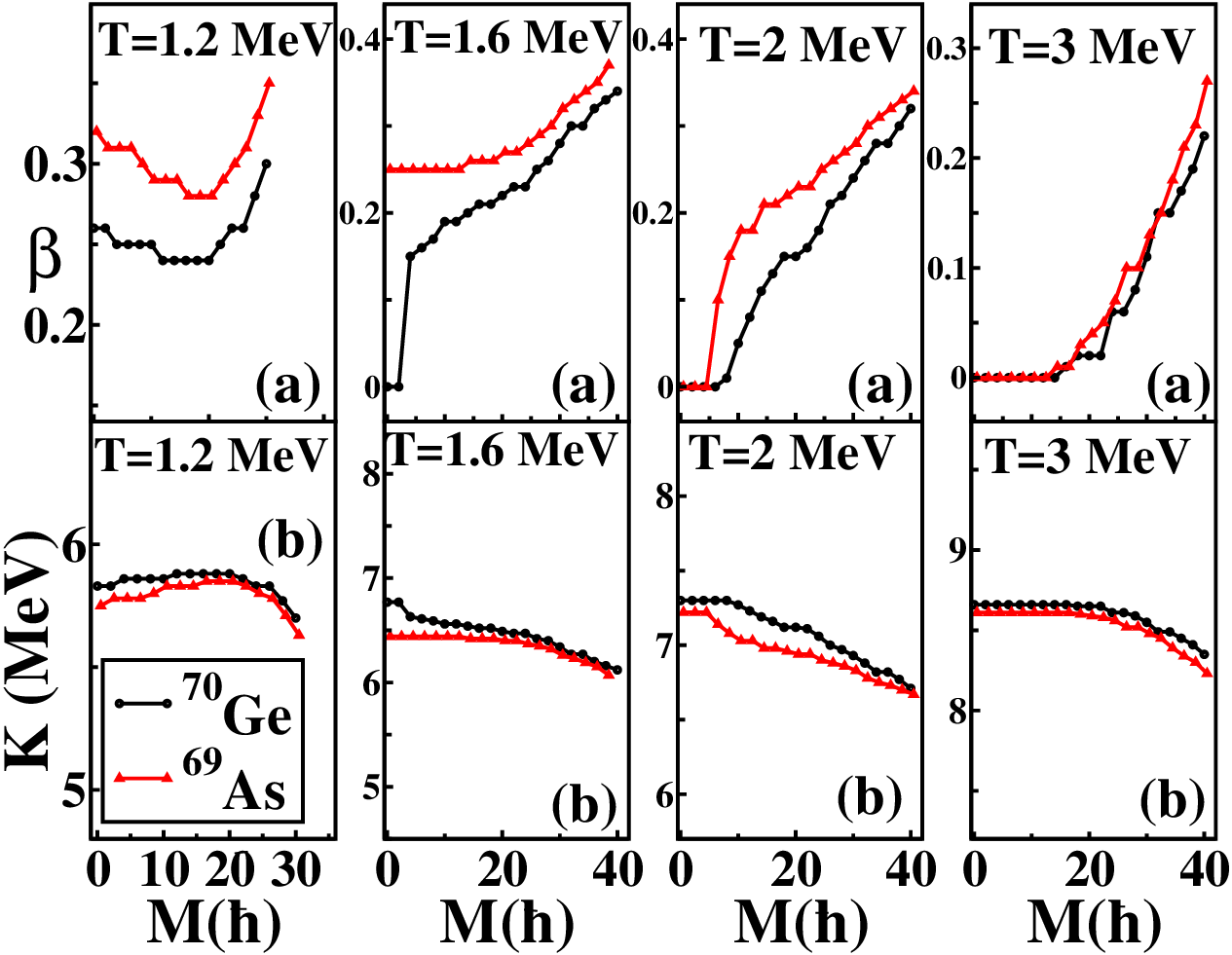}
\caption{(a) deformation ($\beta$) and (b) Inverse Level density parameter 'K'  vs. angular momentum M($\hbar$) for $^{69}$As and $^{70}$Ge at different temperatures.}
\label{kdefvsM}
\end{figure}
The structural properties of any nucleus get altered profoundly once the rotational degree of freedom is incorporated in the calculations. Fig. \ref{kdefvsM} plots the equillibrium deformation and 'K' vs angular momentum  M($\hbar$) for well deformed nuclei $^{69}$As and $^{70}$Ge. The level density is a minima at a shell closure and with increasing deformation the single particle level density of states increases. Several levels from higher oscillator shells which get lowered in energy due to deformation, contribute to the level density that results in the enhancement of LD and decreasing K. In Fig. \ref{kdefvsM}, we find higher $\beta$ and lower 'K' indicating the enhanced LD for a more deformed system $^{69}$As as compared to the lesser deformed $^{70}$Ge. Few experimental evidences ~\cite{KAUSHPLB,JUNGHAN} of the collective enhancement of level density (CELD) and its fadeout with excitation energy have also been reported in recent times. The thermal excitation energy available to a hot rotating system is reduced by increasing the rotational energy at higher spins which results in increasing K indicating a relative reduction of the level density with increasing angular momentum as observed in our calculations in Ref. ~\cite{MPROY} that had shown a good agreement with our evaluated 'K' values. A reasonable agreement of the 'K' variation with temperature and angular momentum with the experimental data ~\cite{MPROY} has instilled further confidence in the reliability of our calculations.\par
In Fig. \ref{kdefvsM}, $\beta$ shows fluctuations at lower and mid-spin values but continue to rise with M($\hbar$) $>$ 20$\hbar$. With increasing deformation at large M values, the single particle level density of states increases and  K decreases (see Fig. \ref{kdefvsM}). Excited states show large deformation  even at T $\approx$ 2 MeV due to rotation at high M($\hbar$) values. However, with further increasing T and the excitation energy, the rise in deformation reduces, the level density enhancement fades out and the 'K' remains almost constant with very minor decline in it's values.\par
\begin{figure}[htb]
\centering
\includegraphics[width=0.42\textwidth]{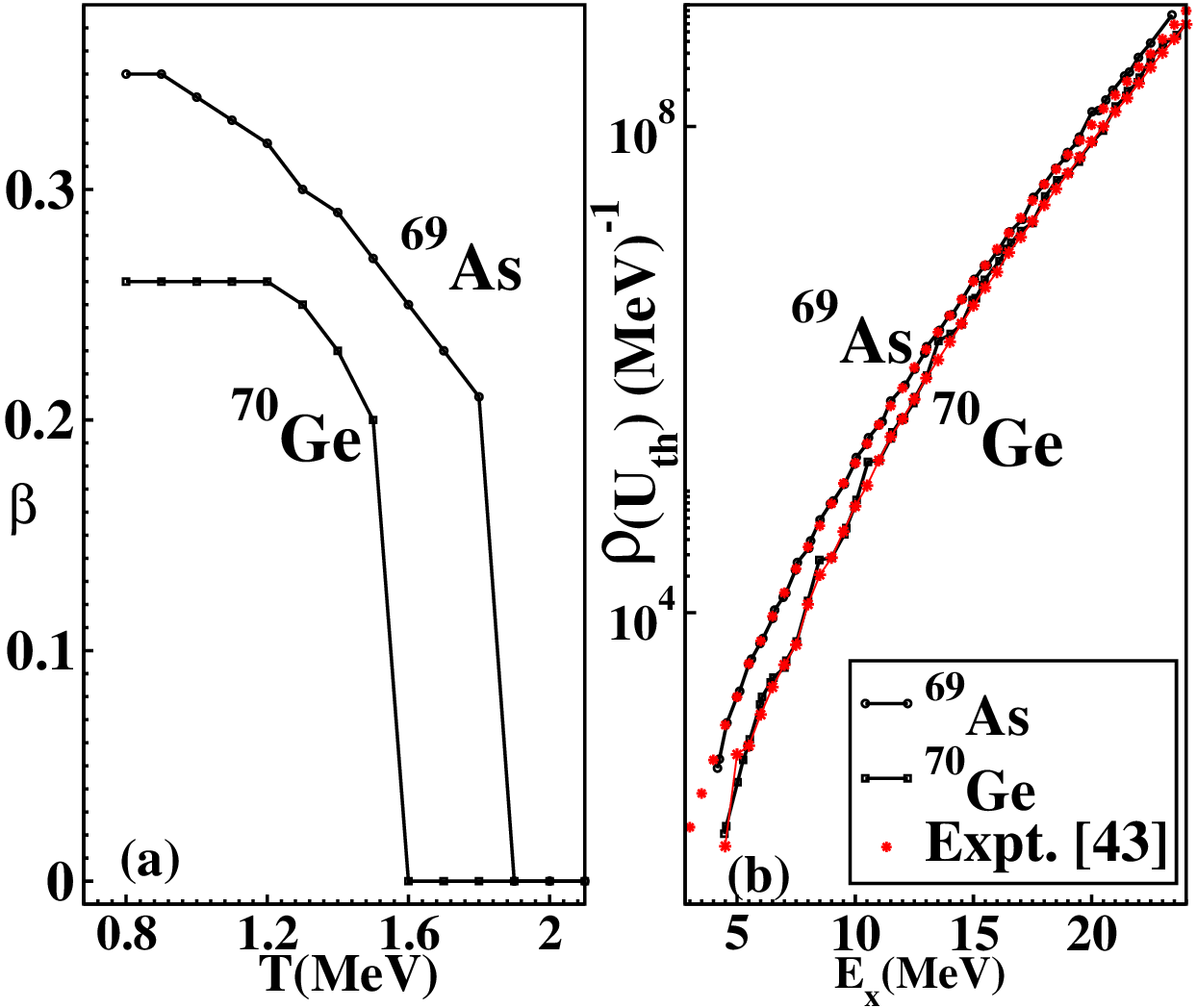} 
\caption{(a) $\beta$ for T corresponding to the excitation energies used for calculating NLD in  ~\cite{UKPAL} (b) Our computed NLD ($\rho$(E$_x$) for $^{69}$As and $^{70}$Ge for the excitation energies given in Ref.~\cite{UKPAL} along with the $\rho$(E$_x$) data from Ref. ~\cite{UKPAL} which show excellent match.}
\label{ldexp}
\end{figure}
The shell structure in the nuclear single-particle levels influences the nuclear level density and the NLD parameter related to the density of single particle levels near the Fermi surface is influenced by the shell structure. A level density parameter 'a' was proposed by Ignatyuk et al. ~\cite{IGNATYUK} which includes an intrinsic excitation energy (E$^*$) dependent shell effect term.  This form of LD parameter exhibits shell effects at low excitation energies and goes over to its asymptotic value at high excitation energies ~\cite{Tatha} exhibiting the phenomenological damping as introduced by Ignatyuk et.al where the shell correction decreases as excitation energy increases. However, unlike using an excitation energy dependent shell correction term exhibiting phenomenological damping as introduced by Ignatyuk ~\cite{IGNATYUK}, and used in Ref. \cite{Tatha} and by Rajasekaran et. al. \cite{MTN}, we incorporate the ground state shell correction along with the deformation effects to the excitation energy, where part of the excitation energy is being used up in overcoming the shell effects and the remaining excitation energy is actually the effective excitation energy [75] available to the nuclear system and is used for all our calculations. This method helps us to evaluate all the statistical quantities with much ease and has been very effective in exhibiting the strong shell effects at low excitation energy which have been shown to disappear with increasing excitation energy in our earlier works ~\cite{MPROY,manpa,MAPLB,MAIJMP4,MA}. Our formalism has been able to estimate level densities and emission probability while interpreting, explaining and reproducing the data effectively as seen earlier ~\cite{MPROY,manpa,MAPLB} .\par

Fig. \ref{ldexp} (a) shows the equillibrium deformation for odd$-$Z $^{69}$As and neighbouring even -even $^{70}$Ge nuclei where $^{69}$As is much more deformed than the even$-$even nucleus $^{70}$Ge. With increasing excitation energy, the shell effects melt away and deformation of both the nuclei drops to zero and the equillibrium shape turns to spherical. However, $^{69}$As remains deformed upto a higher temperature whereas $^{70}$Ge attains zero deformation at much lower T. Since our calculations of excitation energy and total energy of the excited nuclei involve the microscopic corrections including the shell and deformation effects along with an odd$-$even term in the macroscopic mass formula to account for pairing, the shell effects with odd and even nuclei are evidently seen in a very natural way with our formalism. Our model has been very effective in depicting the structural transitions in an easy manner. The microscopic pairing corrections are not included in our calculation as pairing is expected to become negligible at the excitations energy range used in this work. \par

Fig. \ref{ldexp} (b) shows the plots of nuclear level density as a function of excitation energy of the residual nuclei $^{69}$As and $^{70}$Ge. The range of excitation energy data from Ref. ~\cite{UKPAL} is varying from 3 to 24 MeV which has been reproduced using our formalism for T$=$ 0.9 to 2.0 MeV by incorporating microscopic corrections and our calculated proton separation energy ~\cite{MASHA} which is evaluated in similar way as S$_N$$^{cor}$ for protons. We have also plotted the NLD data from the Ref. ~\cite{UKPAL} which is in very good match with the trend of variation and the range of our calculated NLD values. At low T, the NLD curves are apart from each other where the shell and deformation effects are prevalent, but at higher excitation energy the NLD curves almost converge showing the quenching of shell and deformation effects at higher energy. Here pairing may also have some contribution but that has not been included in the calculations. The convergence of NLD seen at high energy is clearly an effect of the damping of shell effects evident in Fig.  \ref{ldexp} (a) and (b). The agreement of data with our computed $\rho$ values gives us the required impetus to extend our calculations to other nuclear systems. Here it is evident that inspite of certain limitations, our formalism has been able to show the shell, deformation effects and structural transitions with excitations and their influence on NLD very efficiently.\par  
\begin{figure}[htb]
\centering
\includegraphics[width=0.42\textwidth]{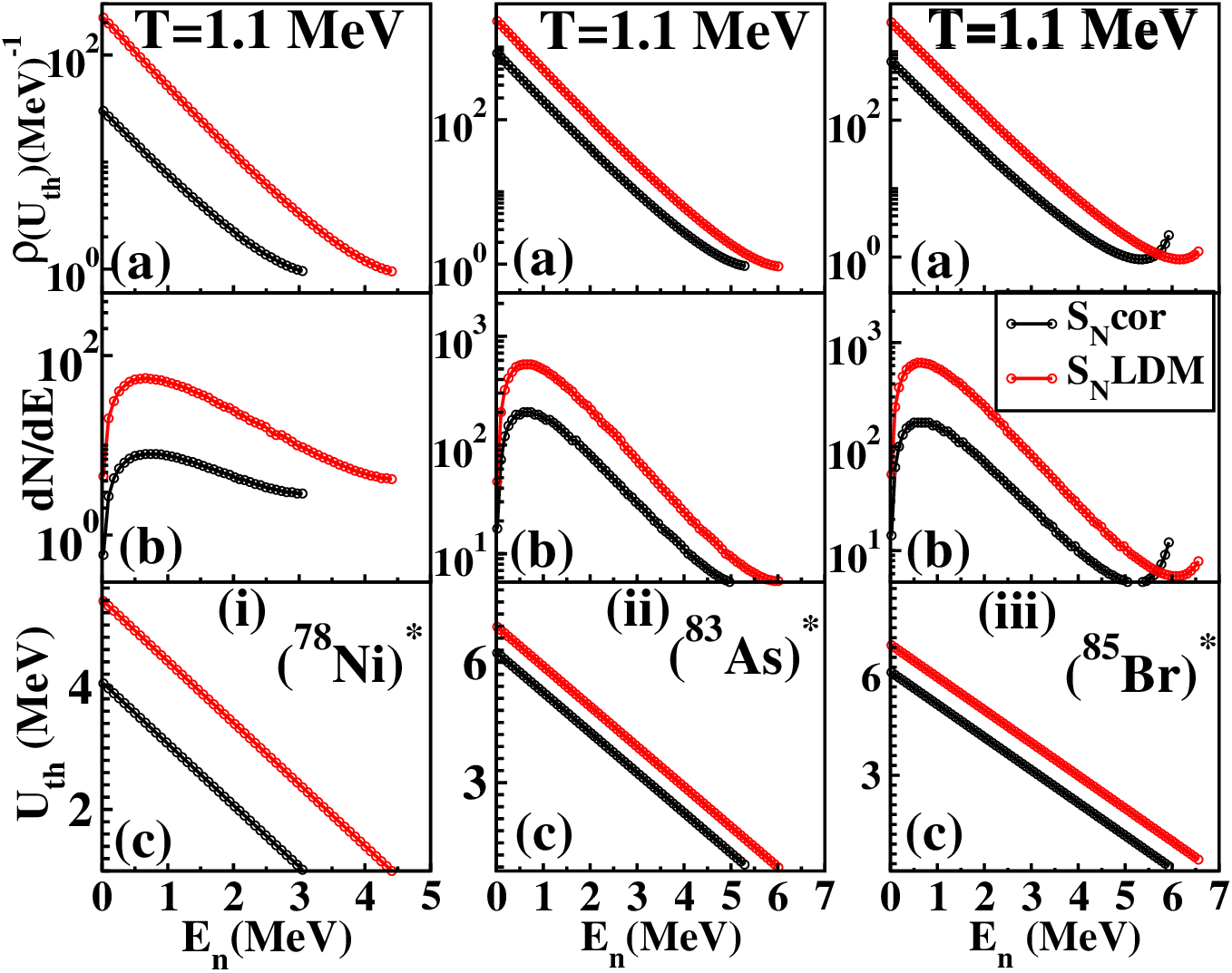}
	\caption{plot of (a) NLD ($\rho$(U$_{th}$) (b) \textit{dN/dE} (Counts/MeV) (from Eq.(\ref{yld})) and (c) U$_{th}$ vs the energy of outgoing particle E$_n$ for (i) $^{78}$Ni$^*$ (Z=28), (ii) $^{83}$As$^*$ (Z$=$33) and (iii) $^{85}$Br$^*$ (Z$=$35) with N$=$50. Two curves in each panel correspond to calculations using  S$_N$$^{cor}$ and  S$_N$$^{LDM}$ respectively.}
\label{ldcorldm}
\end{figure}
Fig. \ref{ldcorldm} plots (a) $\rho$(U$_{th}$) (b) number of neutron emited within the energy interval E$_n$ and E$_n$ +dE$_n$ as \textit{dN/dE} (evaluated using Eq.(\ref{yld})) and (c) U$_{th}$,  as a function of the energy of outgoing particle E$_n$, for the deexciting (i) doubly magic nucleus  $^{78}$Ni$^*$ (Z=28), mid-shell nuclei (ii) $^{83}$As$^*$ (Z$=$33) and (iii) $^{85}$Br$^*$ (Z$=$35) with magic neutron number N$=$50. The two curves in each panel denoted by S$_N$$^{cor}$ and  S$_N$$^{LDM}$ correspond to the calculations of U$_{th}$, NLD and  \textit{dN/dE} by using the neutron separation energy with and without the inclusion of shell correction and deformation effects. All the curves for U$_{th}$, NLD and  \textit{dN/dE} show the impact of inclusion of shell correction. This calculation has been done at a temperature smaller than the critical temperature so that the shell effects are not washed out. Although all the nuclei shown are magic with N=50, the extra stability of the doubly magic nucleus leads to a lower residual excitation energy as the larger part of the excitation energy is used up to overcome the shell effects which is clearly seen in the lower U$_{th}$ and consequently in lower NLD and yield in the residual nucleus ($^{77}$Ni) of doubly magic $^{78}$Ni$^*$ than that of $^{83}$As$^*$ and $^{85}$Br$^*$. Secondly, the large gap between the curves of U$_{th}$, NLD and Yield corresponding to  S$_N$$^{cor}$ and  S$_N$$^{LDM}$ shows the impact of inclusion of shell correction. This gap is maximum in doubly magic $^{78}$Ni$^*$ (Fig (i)(a),(b),(c)) as compared to $^{83}$As$^*$ (Fig (ii)(a),(b),(c)) and $^{85}$Br$^*$ (Fig (iii)(a),(b),(c)) showing the highest shell correction in doubly magic nucleus. The level density and the emission spectra is minimum in doubly magic  $^{78}$Ni$^*$. The enhanced LD in the case of $^{83}$As$^*$ and $^{85}$Br$^*$ as compared to doubly magic $^{78}$Ni$^*$ reflects the impact of shell effects. This emphasizes the importance of including shell correction in the evaluation of NLD ~\cite{PCROUT}. \par

\begin{figure}[htb]
\centering
\includegraphics[width=0.42\textwidth]{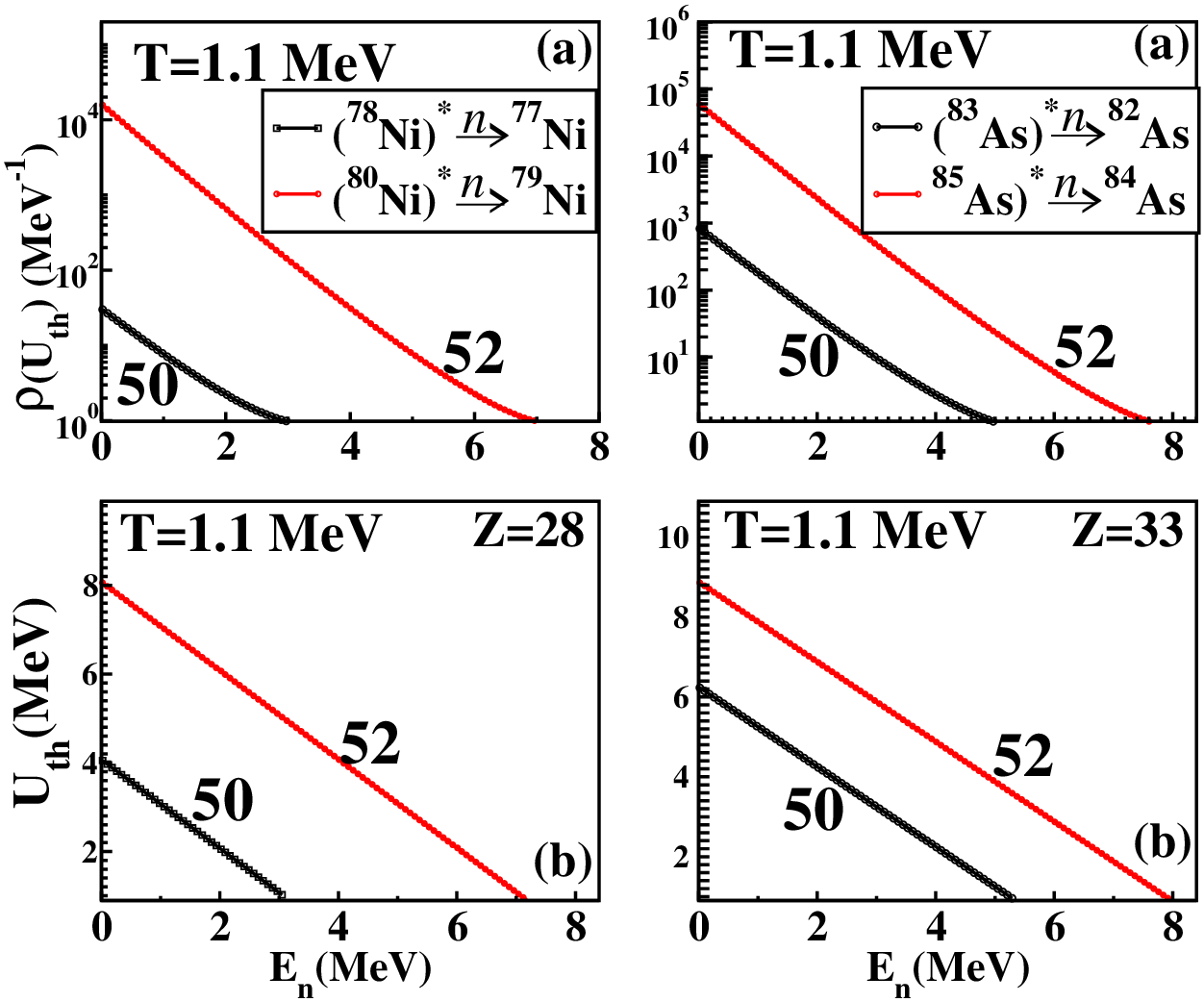}
\caption{(a) NLD and (b) U$_{th}$ for the residual nuclei (i) $^{77}$Ni (ii)$^{79}$Ni (iii)$^{82}$As and (iv) $^{84}$As after 1 neutron emmision from ((i) $^{78}$Ni$^*$ and $^{80}$Ni$^*$ (ii) $^{83}$As$^*$ and $^{85}$As$^*$) respectively at T$=$1.1 MeV. The numbers 50 and 52 on the curves represent neutron number of parent nuclei}
\label{ld5052}
\end{figure}
\begin{figure}[htb]
\centering
\includegraphics[width=0.42\textwidth]{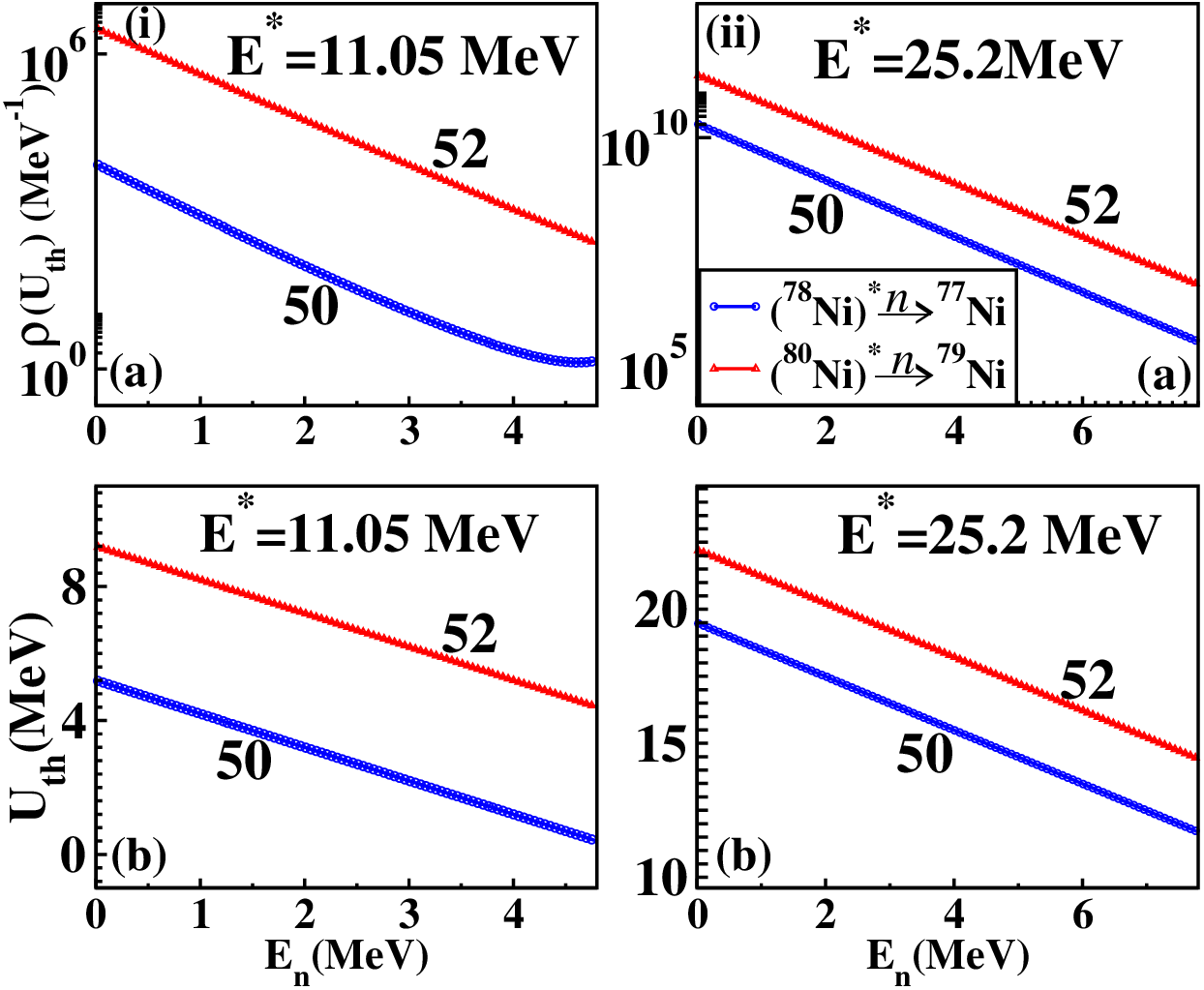}
\caption{(a) NLD and (b) U$_{th}$ for the residual nuclei $^{77}$Ni and $^{79}$Ni after 1 neutron emmision from $^{78}$Ni$^*$ (N$=$50) and $^{80}$Ni$^*$ (N$=$52) respectively at Ex$*$ $=$ (i) 11.05 MeV and (ii) 25.2 MeV. The difference in Separation energy of Ni isotopes (N$=$50 and 52) is reflected in their respective U$_{th}$ and NLD curves. The numbers 50 and 52 on the curves represent neutron number of parent nuclei}
\label{ldex5052}
\end{figure}
\vspace{0.5cm}

Fig. \ref{ld5052} shows the  U$_{th}$ and NLD for neutron evaporation from the excited compound nuclei (i) $^{78}$Ni$^*$ and $^{80}$Ni$^*$ (ii) $^{83}$As$^*$ and $^{85}$As$^*$. Here we have chosen N$=$50 and 52 isotopes of the parent nuclei with Z$=$28 and Z$=$33 for evaluating NLD at a fixed temperature T$=$1.1 MeV. The sudden drop in separation energy by around $\approx$ 3 MeV in nucleus (with N$=$52) just after the magic neutron number (N$=$50) (as seen in Fig. \ref{gsn} for many nuclei) results in much lower U$_{th}$ (from Eq. \ref{ures}). After 1n-emission from $^{78}$Ni$^*$, the residual nucleus $^{77}$Ni has much lower  U$_{th}$ than in residual nucleus ($^{79}$Ni) of $^{80}$Ni$^*$. Lower U$_{th}$ results in lower NLD as is visible in Fig. \ref{ld5052}. Also, the decline in S$_n$ value in going from neutron number 50 to 52 is much different in closed shell $_{28}$Ni$_{50}$ and mid-shell $_{33}$As$_{50}$ nuclei. This is reflected in the large gap between the corresponding curves of  U$_{th}$ and NLD in $^{77}$Ni and $^{79}$Ni as compared to that in $^{82}$As and $^{84}$As,  pointing towards the extra stability of closed shell than mid shell nuclei. Fig \ref{ldex5052} displays the influence on NLD variation due to separation energy dropping from N=50 to 52 at excitation energies Ex$^*$ =  11.05 MeV and 25.2 MeV. Plots of U$_{th}$ and NLD for residual nuclei $^{77}$Ni and $^{79}$Ni show that with increasing excitation energy,  U$_{th}$ and NLD increase and the shell effects quench significantly which is evident in the two curves representing $^{77}$Ni and $^{79}$Ni coming closer to each other indicating the damping of shell effects. \par 

\vspace{0.5cm}
\begin{figure}[htb]
\centering
\includegraphics[width=0.42\textwidth]{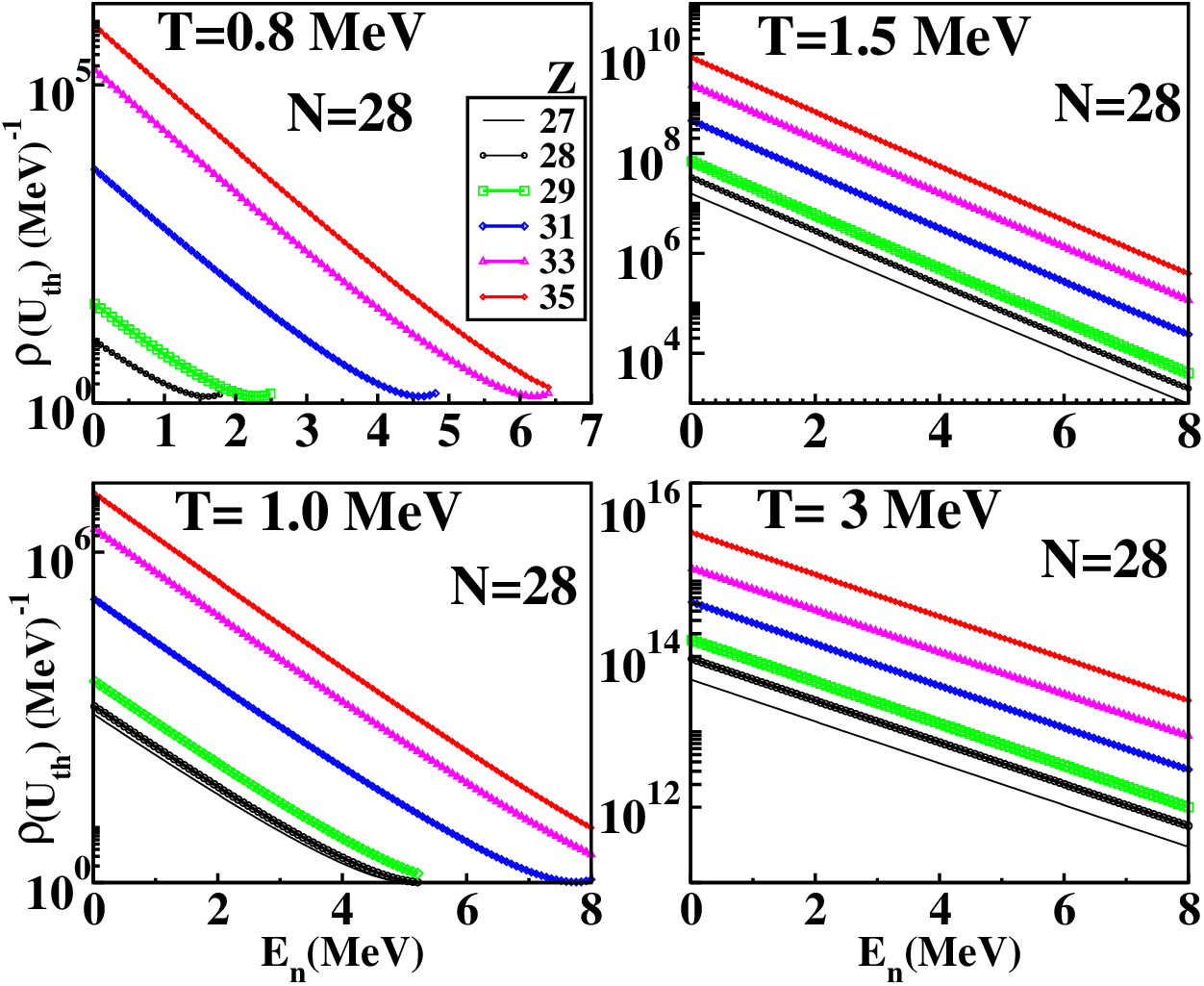}
\caption{ NLD ($\rho$(U$_{th}$) for 1-n emission from the N=28 isotones of Z $=$ 28 and odd-Z $=$ 27$-$ 35 for T=0.8, 1.0, 1.5 and 3.0 MeV exhibiting shell effects and the quenching of shell effects.}
\label{ldexi}
\end{figure}
Fig. \ref{ldexi} shows the nuclear level density vs. En for T$=$ 0.8, 1.0, 1.5 and 3.0 MeV for the deexciting compound nuclei  $^{55}$Co$^*$, $^{56}$Ni$^*$, $^{57}$Cu$^*$,$^{59}$Ga$^*$, $^{61}$As$^*$ and $^{63}$Br$^*$ (N$=$28 isotones of Z $=$ 27$-$ 35). $^{56}$Ni$^*$ being doubly magic is spherical in shape with  zero deformation. The neighbouring nuclei Z$=$ 27 and 29 also show very small deformation. While moving away from the shell closure (Z$=$28), the deformation increases which can be seen in the enhanced level density and closely spaced curves for well deformed nuclei (with Z$=$31, 33 and 35) at low temperatures (T $=$0.8 MeV) where the  shell effects are predominant. Here it is important to mention that although the nuclei $^{59}$Ga$^*$, $^{61}$As$^*$ and $^{63}$Br$^*$ are magic with N$=$28, but the magic character of these neutron deficient nuclei was seen to be weakening as mentioned in the earlier discussion. These nuclei are well deformed that contributes to the enhancement of level density. As the excitation energy increases, the gap between the curves reduces. At T$=$ 1.5 to 3 MeV, all the curves look equidistant and close to each other varying smoothly indicating quenching of shell effects. The level density increases with increasing excitation energy as expected for all nuclei. But a signficant difference that exists between the NLD of nuclei at or near shell closure and mid-shell nuclei at low T, eventually starts disappearing at high excitation energy. This shows the impact of shell effects as well as the quenching of shell effects with excitation energy on NLD shown in a very simple yet effective manner. \par 
\vspace{0.5cm}
\begin{figure}[htb]
\centering
\includegraphics[width=0.42\textwidth]{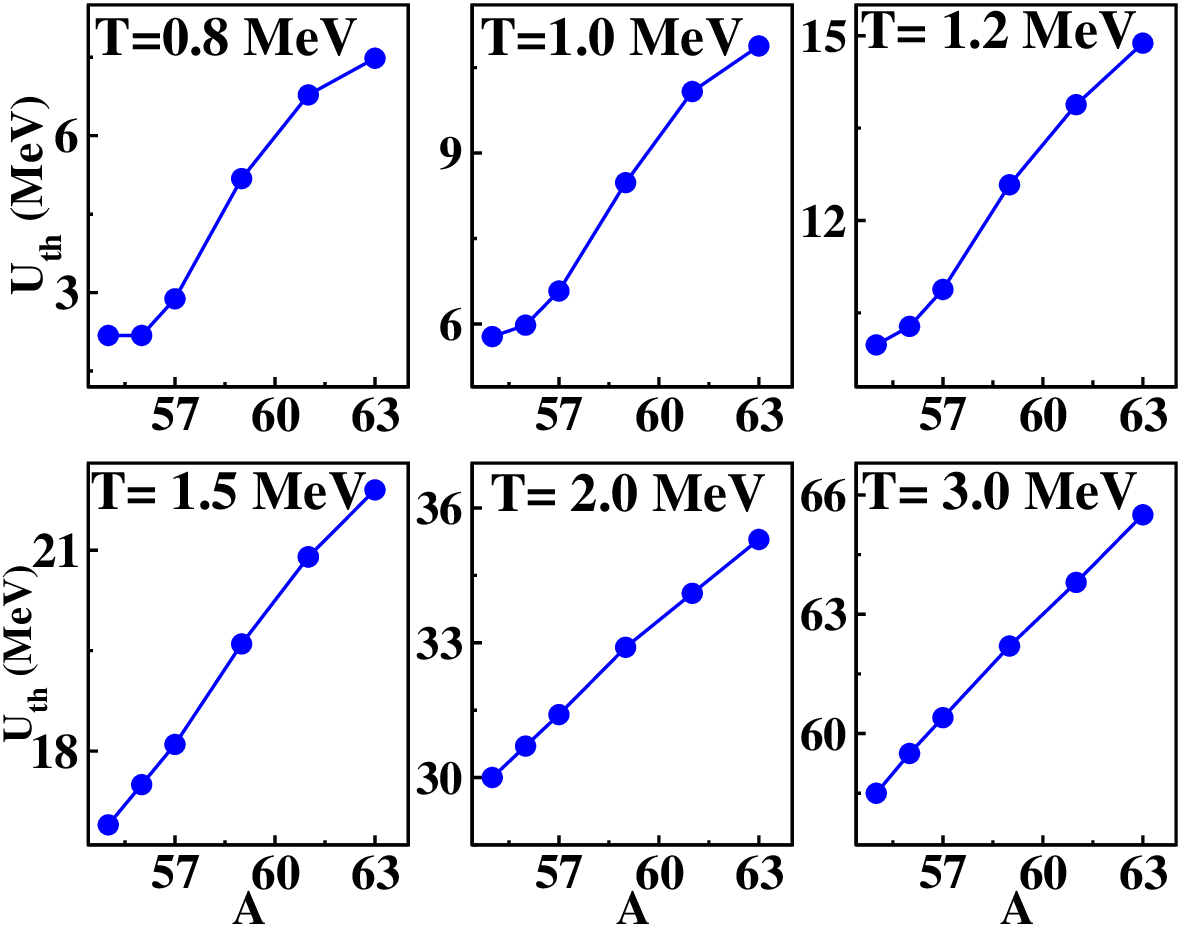}
\caption{U$_{th}$ for N=28 isotones of odd-Z $=$ 27$-$ 35 for T=0.8, 1.0, 1.5 and 3.0 MeV showing the variation of excitation energy with T}
\label{uthall}
\end{figure}
The excitation energy evaluated for all these nuclei is plotted in Fig \ref{uthall}. Since T is the input in our calculations, we show the variation of excitation energy vs. T that exhibits the shell effects at very low T. \par

\begin{figure}[htb]
\centering
\includegraphics[width=0.42\textwidth]{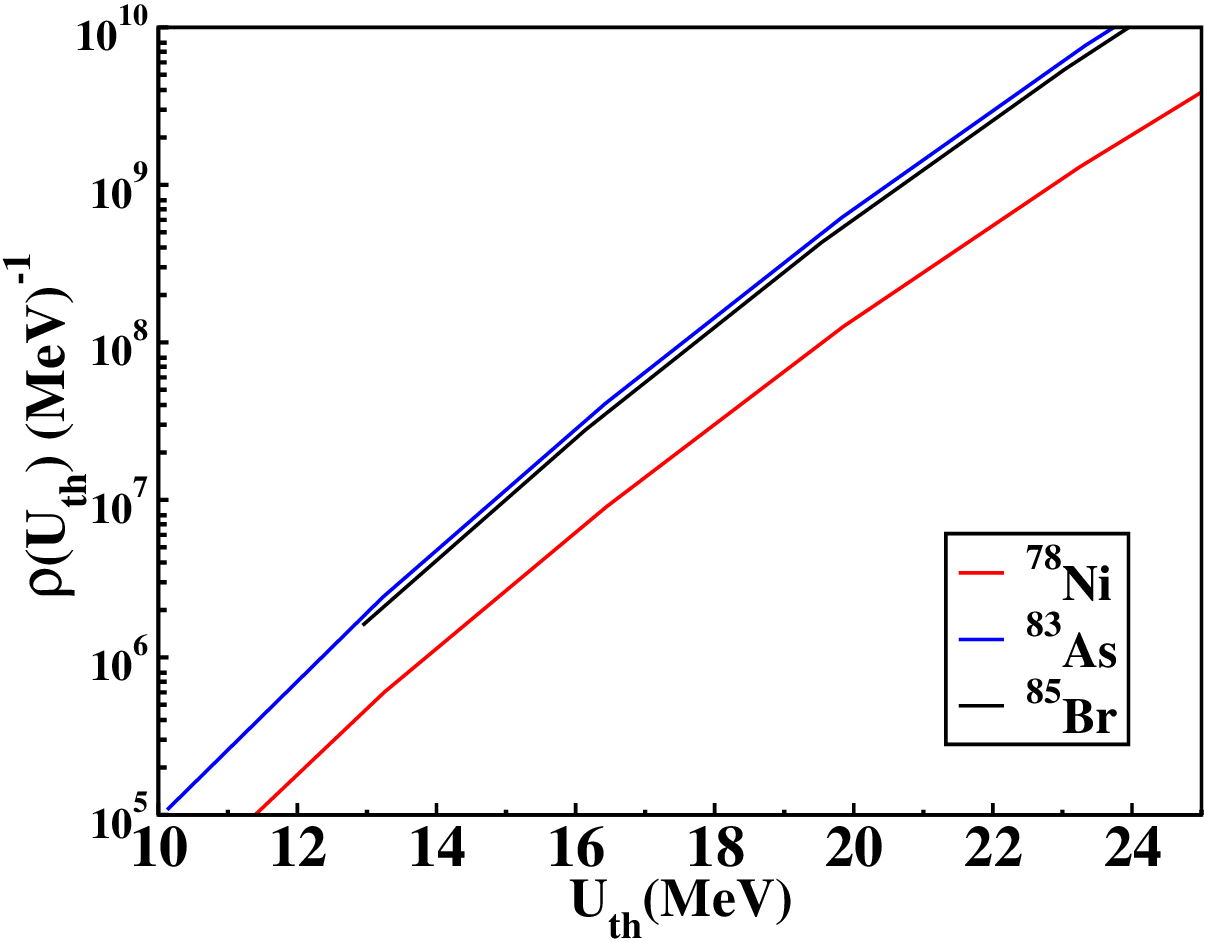}
\caption{Variation of NLD ($\rho$(U$_{th}$)) vs U$_{th}$ $^{78}$Ni$^*$, $^{83}$As$^*$, $^{85}$Br$^*$.}  
\label{lduth}
\end{figure}
Fig. \ref{lduth} shows the variation of NLD as a function of excitation energy U$_{th}$ for magic nucleus $^{78}$Ni$^*$ and non-magic nuclei  $^{83}$As$^*$, $^{85}$Br$^*$. NLD is much less for a magic nucleus as compared to well deformed non magic nuclei.\par

\begin{figure}[htb]
\centering
\includegraphics[width=0.42\textwidth]{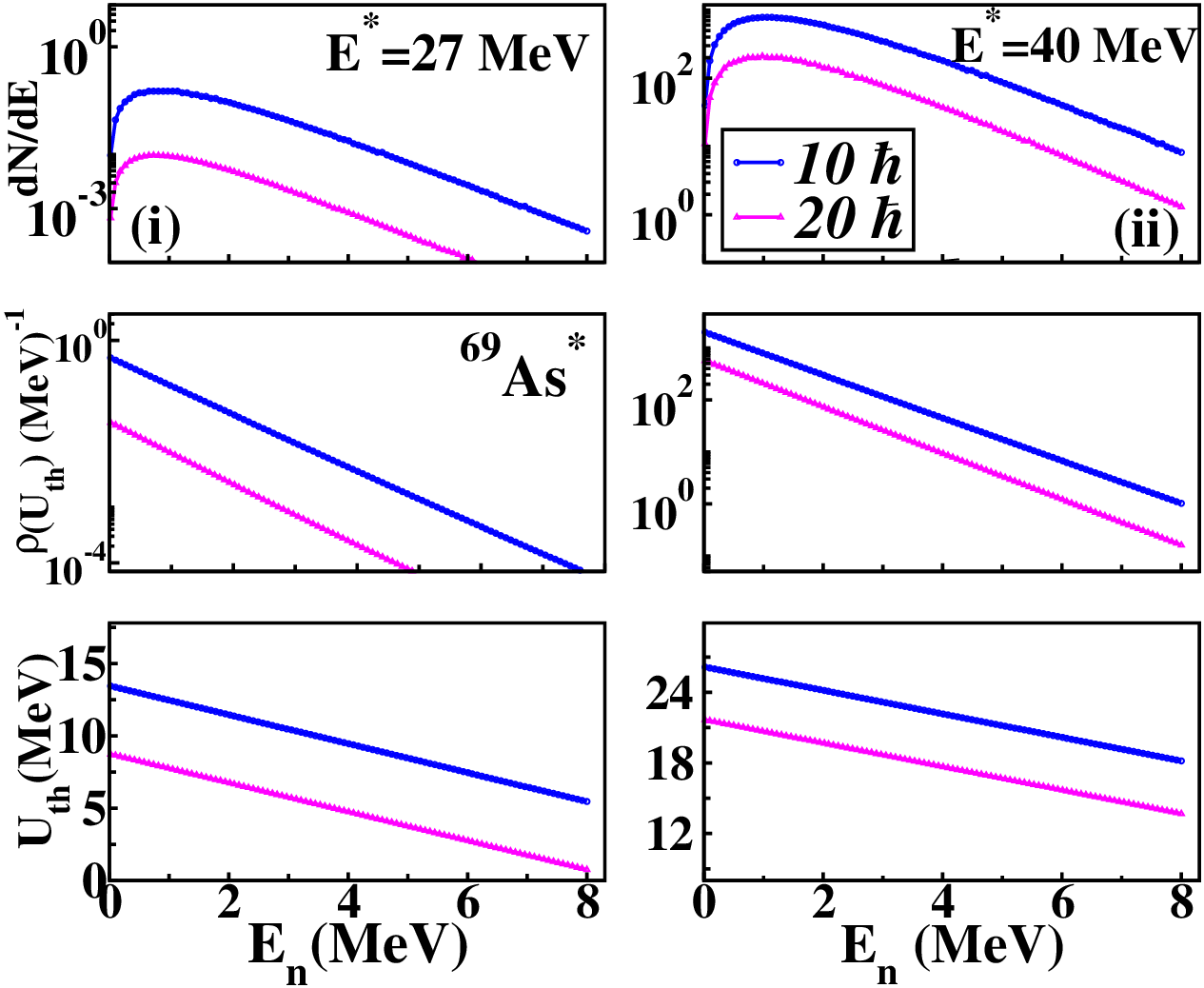}
\caption{U$_{th}$, NLD and  \textit{dN/dE} (Counts/MeV) from Eq.(\ref{yld}) for $^{69}$As$^*$  at E$^*$ $=$ (i) 27 MeV and (ii) 40 MeV at M($\hbar$)= 10, 20.}
\label{ldmh}
\end{figure}

The collective enhancement of LD and it's fadeout is shown in Fig. \ref{ldmh} where we have plotted U$_{th}$, $\rho$(U$_{th}$) and dN/dE to get yield (P), vs. E$_n$ for excitation energy E$^*$ $=$ 27 MeV and 40 MeV and angular momentum 10$\hbar$ and 20$\hbar$ for a well deformed nucleus $^{69}$As$^*$.  As excitation energy increases, U$_{th}$ and consequently $\rho$(U$_{th}$) and number of neutron emitted dN/dE increase significantly.  As angular momentum increases, the rotational energy increases and the thermal energy available to the residual nucleus decreases which reduces the level density and neutron emission probability which is evident from Fig. \ref{ldmh} where U$_{th}$, $\rho$(U$_{th}$) and  \textit{dN/dE} are shown. The reducing gap between the curves tending to the convergence of curves with increasing excitation energy indicates the fade out of LD enhancement. \par
\vspace{0.5cm}
\begin{figure}[htb]
\centering
\includegraphics[width=0.46\textwidth]{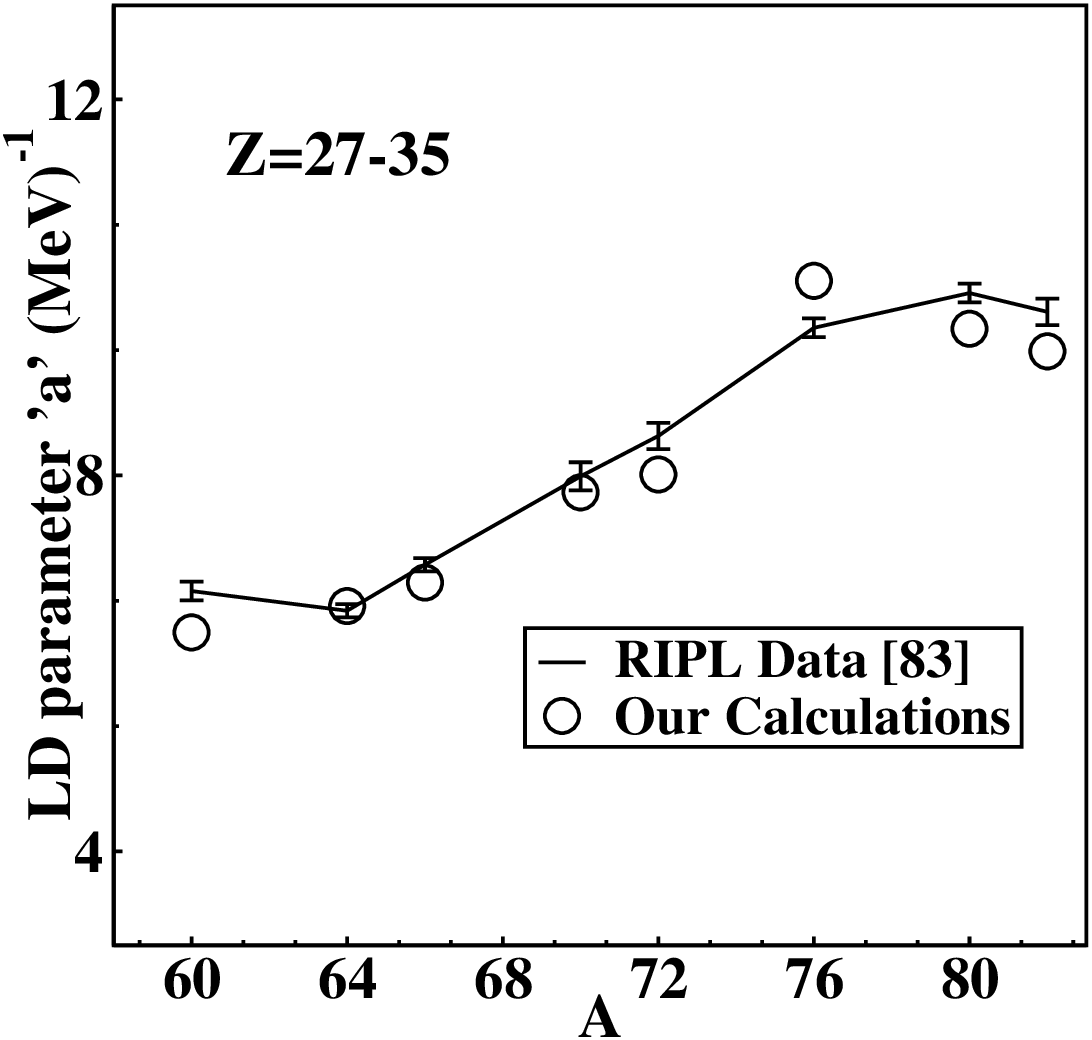}
\caption{Our calculated Level density parameter 'a'(MeV) for Z $=$ 27$-$ 35 compared with LD parameters for BSFG Model taken from RIPL-2 ~\cite{ripl}.}
\label{ldripl}
\end{figure}

Fig.\ref{ldripl} shows the plot of level density parameter 'a' for the BSFG model from RIPL-2 (Reference Input Parameter Library) ~\cite{ripl} and the 'a' values for the corresponding nuclei evaluated by our theoretical calculations. These level density parameters for the BSFG model have been obtained by fitting the Fermi-gas model formula to the recommended spacings of s-wave neutron resonances and to the cumulative number of low-lying levels evaluated from the analysis of nuclear levels provided by Ignatyuk et al ~\cite{IGNAT}. Our calculated values of LD parameter 'a' show a reasonable agreement in the overall trend and range of 'a' values with that obtained from the resonance spacing analysis. Our objective to see a close match in trend and the range of variation of 'a' with the fitted data has been accomplished although some of the values are not in close agreement. The availability of new experimental data of NLD will help a better comparison, analysis and interpretation of data by more detailed investigations using the theoretical prescriptions like ours and many other works. Some more rigorous theoretical and experimental work is needed in this direction to have more clarity on this subject. \par

\section{Conclusion}
To conclude, a microscopic calculation has been performed within the framework of microscopic Statistical Model to investigate the impact of shell effects on the nuclear level density and consequently the particle evaporation spectra for nuclei from  Z$=$ 27 to 35. The quenching of shell effects with increasing excitation energy and their impact on NLD has been studied. The separation energy which exhibits shell effects plays an important role in NLD variation. The importance of including the microscopic corrections due to shell effects while evaluating NLD has been highlighted in much easy and effective way. The closed shell and mid-shell nuclei respond to excitation differently which is reflected in NLD. Inverse level density parameter exhibits strong shell effects as well as the washing out of shell effects with increasing excitation energy. The enhancement of LD parameter with the deformation and rotation with the fade out of enhancement with increasing excitations has been shown. Excited nuclei near the proton drip line exhibit the weakening of magicity of N$=$28 observed in the inverse level density parameter variation. Our calculated level density parameter agrees well with RIPL-2 data. Our calculations of NLD have been able to reproduce the available data which shows the reliability of our theoretical formalism.   \par

\section*{Acknowledgements}
Author acknowledges the financial support by Science and Engineering Research Board (DST), Govt. of India under the WOS-A scheme. \par

\end{small}

\begin{thebibliography}{99}
\bibitem{BETHE} H. Bethe, \textit{Phys. Rev.} \textbf{50} (1936) 332; \textit{Rev. Mod. Phys.} \textbf{9} (1937) 69.
\bibitem{ERIC} T. Ericson, \textit{Adv. Phys.} \textbf{9} (1960) 425.
\bibitem{BOHR} A. Bohr and B. R. Mottelson, in Nuclear Structure (Benjamin, New York, 1969), Vol. I, P 281.
\bibitem{IGNATYUK} A. V. Ignatyuk, G. N. Smyrenkin, M. G. Itkis, S. I. Mulgin, and V. N. Okolovich, Sov. J. Part. Nuclei \textbf{16}, 307 (1985), Fiz. Elem. Chastits At. Yadra 16, 710 (1985).
\bibitem{KATARIA} S. K. Kataria and V. S. Ramamurthy, Nucl. Phys. \textbf{A 349} (1980) 10.
\bibitem{MTN} M. Rajasekaran T. R. Rajasekaran, N. Arunachalam, Phys. Rev C \textbf{37} (1988) 307.
\bibitem{WALLA} R. K. Wallace and S. E. Woosely, Astrophys. J. Suppl. \textbf{45} (1981) 389.
\bibitem{KAUSHIK} K. Banerjee, et. al.,  Phys. Rev. \textbf{C 85} (2012) 064310.
\bibitem{BALARAM} Balaram Dey et. al.,Phys. Rev. \textbf{C 91} (2015) 044326.
\bibitem{GOHIL} M. Gohil et. al., Phys. Rev. C 91 (2015) 014609; EPJ Web of Conf. \textbf{66} (2014) 03073.
\bibitem{YKGUP} Y. K. Gupta, B. John, D. C. Biswas, B. K. Nayak, A. Saxena, and R. K. Choudhury, Phys. Rev. \textbf{C 78} (2008) 054609.
\bibitem{PROY} P. Roy, et. al., Phys. Rev. C 86 (2012) 044622.
\bibitem{PCROUT} P. C. Rout et. al., Phys. Rev. Lett. \textbf{110} (2013) 062501.
\bibitem{Schiller} A. Schiller, A. Bjerve, M. Guttormsen, M. Hjorth-Jensen, F. Ingebretsen, E. Melby, S. Messelt, J. Rekstad, S. Siem, and S. W. Odegard, Phys. Rev. \textbf{C 63}, 021306(R) (2001).
\bibitem{Guttormsen} M. Guttormsen, E. Melby, J. Rekstad, S. Siem, A. Schiller, T.Lonnroth, and A. Voinov, J. Phys. \textbf{G 29}, 263 (2003).
\bibitem{Melby} E. Melby et al., Phys. Rev. Lett. \textbf{83}, 3150 (1999).
\bibitem{SCHILLER1} A. Schiller, L. Bergholt, M. Guttormsen, E. Melby, J. Rekstad, and S. Siem, Nucl. Instrum. Methods Phys. Res. \textbf{A 447}, 498 (2000).
\bibitem{KATSANOS} A. A. Katsanos, J. R. Huizenga, and H. K. Vonach, Phys. Rev.
\textbf{141}, 1053 (1966).
\bibitem {Katsanos1} A. A. Katsanos and J. R. Huizenga, Phys. Rev. \textbf{159}, 931 (1967).
\bibitem{Raman} S. Raman, T. A. Walkiewicz, S. Kahane et al., Phys. Rev. \textbf{C 43}, 521 (1991).
\bibitem{Mishra} V. Mishra, N. Boukharouba, C. E. Brient, S. M. Grimes, and R. S. Pedroni, Phys. Rev. \textbf{C 49}, 750 (1994).
\bibitem{Capote} R. Capote, M. Herman, P. Oblozinsky et al., Nucl. Data Sheets \textbf{110}, 3107 (2009).
\bibitem{Hasse} R. W. Hasse and P. Schuck, Phys. Lett. \textbf{B 179} (1986) 313.
\bibitem{Nebbia} G. Nebbia, D. Fabris, A. Perin et al., Nucl. Phys. \textbf{A 578}, 285 (1994).
\bibitem{Hagel} K. Hagel, D. Fabris, P. Gonthier et al., Nucl. Phys. \textbf{A 486}, 429 (1988).
\bibitem{wada} R. Wada, D. Fabris, K. Hagel et al., Phys Rev. \textbf{C 39}, 497 (1989).
\bibitem{Gonin} M. Gonin, L. Cooke, K. Hagel et al., Phys. Lett. \textbf{B 217}, 406 (1989).
\bibitem{Gonin1} M. Gonin, L. Cooke, K. Hagel et al., Phys. Rev. \textbf{C 42}, 2125 (1990).
\bibitem{Chbihi} A. Chbihi, L. G. Sobotka, N. G. Nicolis, D. G. Sarantites, D. W. Stracener, Z. Majka, D. C. Hensley, J. R. Beene, and M. L. Halbert, Phys Rev. \textbf{C 43}, 666 (1991).
\bibitem{Yoshida} K. Yoshida, J. Kasagi, H. Hama, M. Sakurai, M. Kodama, K.Furutaka, K. Ieki, W. Galster, T. Kubo, M. Ishihara, and A. Galonsky, Phys. Rev. \textbf{C 46}, 961 (1992).
\bibitem{Fabris} D. Fabris, E. Fioretto, G. Viesti et al., Phys. Rev. C \textbf{50}, 1261(R) (1994).
\bibitem{Nebbia1} G. Nebbia, K. Hagel, D. Fabris et al., Phys. Lett. B \textbf{176}, 20 (1986).
\bibitem{Fineman} B. J. Fineman, K. T. Brinkmann, A. L. Caraley, N. Gan, R. L. McGrath, and J. Velkovska, Phys. Rev. C \textbf{50}, 1991 (1994).
\bibitem{Caraley} A. L. Caraley, B. P. Henry, J. P. Lestone, and R. Vandenbosch, Phys. Rev. \textbf{C 62}, 054612 (2000).
\bibitem{Charity1} R. J. Charity, L. G. Sobotka, J. F. Dempsey et al., Phys. Rev. C \textbf{67}, 044611 (2003).
\bibitem{Charity} R. J. Charity, Phys. Rev. C \textbf{82}, 014610 (2010).
\bibitem{BABA} H. Baba, Nucl. Phys. A \textbf{159} (1970) 625.
\bibitem{facchini} U. Facchini and E. Saetta-Menichella, Energia Nucleare \textbf{15} (1968) 54,
\bibitem{HBABA} H. Baba and S. Baba, Japan Atomic Energy Research Institute Report, JAERI \textbf{1183} (1969).
\bibitem{TDNEWTON} T. D. Newton, Can. 3. Phys. \textbf{34} (1956) 804].
\bibitem{manpa} Mamta Aggarwal, Nucl. Phys. A \textbf{983}, (2019) 166.
\bibitem{MAK} Mamta Aggarwal and S. Kailas, \textit{Phys. Rev. C} \textbf{81} (2010) 047302.
\bibitem{UKPAL} U. K. Pal, D. R. Chakrabarty, V. M. Datar, Suresh Kumar, E. T. Mirgule and H. H. Oza, J. Phys. G: Nucl. Part. Phys. \textbf{25} (1999) 1671.
\bibitem{ACHOUDHARI} A. Choudhari et.al., Phys. Rev. C \textbf{91}, (2015) 044620.
\bibitem{SGORIELY}  S. Goriely, Nucl. Phys. A \textbf{605} (1996) 28-60.
\bibitem{VM} V. M. Strutinsky, Nucl. Phys. A \textbf{95} (1967) 420.
\bibitem{MASHA} Mamta Aggarwal,  Phys. Rev. \textbf{89} (2014) 024325.
\bibitem{MAPLB} Mamta Aggarwal, Phys. Lett. B \textbf{693} (2010) 489.
\bibitem{ogane} Yu. Ts. Oganessian et al., Phys. Rev. Lett. \textbf{104}, 142502 (2010).
\bibitem{HOFMAN} S. Hofmann and G. Munzenberg, Rev. Mod. Phys. \textbf{72}, 733 (2000).
\bibitem{Polikanov} S. M. Polikanov, Usp. Fiz. Nauk \textbf{94}, 43 (1968) Sov. Phys. Usp. 11, 22 (1968).
\bibitem{Janssens} R. V. F. Janssens and T. L. Khoo, Annu. Rev. Nucl. Part. Sci. \textbf{41}, 321 (1991).
\bibitem{YADAV} H. L. Yadav, M. Kaushik, and H. Toki, Int. Jour. Mod. Phys. \textbf{E 13} (2004) 647.
\bibitem{bastin} B. Bastin \textit{et al.}, Phys. Rev. Lett. \textbf{99} (2007) 022503.
\bibitem{GMAIJ17} G. Saxena, M. Kumawat, M. Kaushik, U. K. Singh, S. K Jain, S. Somorendro Singh, and Mamta Aggarwal, Int. Jour. Mod. Phys. \textbf{E 26} (2017) 175007.
\bibitem{MPROY} P. Roy et al., Phys. Rev. C \textbf{103}, 024602 (2021).
\bibitem{MACO} Mamta Aggarwal, Phys. Rev. \textbf{C 90} (2014) 064322.
\bibitem{MAMHIAS} Mamta Aggarwal and S. Kailas, EPJ Web of Conf. \textbf{63} (2013) 01016.
\bibitem{MI} M. Aggarwal and I. Mazumdar, Phys. Rev. C \textbf{80} (2009) 024322.
\bibitem{MAJNP} M. Aggarwal, Journal of Nucl. Phys. Material Sci. Radiation and Applications (JNPMSRA) \textbf{3}, No. 2 (2016) 179. 
\bibitem{IGNAT} A. Ignatyuk et al. Nucl. Phys. A \textbf{346} (1980) 191.
\bibitem{GOOD} A. L. Goodman, Phys. Rev. C 35 (1987) 2338.
\bibitem{ALHA} Y. Alhassid, S. Levit, J. Zingman, Phys, Rev. Lett. \textbf{57} (1986) 539.
\bibitem{KAUSHPLB} K. Banerjee et. al, Phys. Lett. B \textbf{772} (2017) 105--109
\bibitem{MA} Mamta Aggarwal, Phys. Rev. {\bf{C 69}}, 034602 (2004).
\bibitem{MAJNP1} Mamta Aggarwal, J. Nucl. Phy. Mat. Rad. A \textbf{5}, No-2 (2018) 255.
\bibitem{MAIJMP8} M. Aggarwal, Int. J. of Mod. Phys. E \textbf{17} (2008) 1091.
\bibitem{MAIJMP4} M. Aggarwal, Int. J. of Mod. Phys. E \textbf{13} (2004) 1239.
\bibitem{BLAT} J.M. Blatt, V.F. Weisskopf, Theoretical Nuclear Physics, Springer-Verlag, New York, 1979.
\bibitem{HM} J. B. Huizenga and L. G. Moretto, Annu. Rev. Nucl. Sci. {\bf{22}}, 427 (1972).
\bibitem{PM} P. Moller and J. R. Nix, At. Data Nucl. Data Tables \textbf{26} (1981) 165.
\bibitem{MNV} M. Rajasekaran, N. Arunachalam, and V. Devanathan, Phys. Rev. {\bf{C 36}}, 1860 (1987).
\bibitem{MORET}L. G. Moretto, Nucl. Phys. A182 (1972) 641; ibid. A185 (1972) 145; ibid. A216 (1973) 1.
\bibitem{EIS} J. M. Eisenberg and W. Greiner, {\it{Microscopic Theory of Nucleus }}(North Holland, New York, 1976).
\bibitem{GS} G. Shanmugam, P. R. Subramanian, M. Rajasekaran, and V. Devanathan, Nuclear Interactions, Lecture Notes in Physics, {\bf{Vol. 72}}, p433 (Springer, Berlin,1979).
\bibitem{HW} D. L. Hill and J. A. Wheeler, Phys. Rev. {\bf{89}}, 1102 (1953).
\bibitem{BIJY} B. K. Agrawal and A. Ansari, Nucl. Phys. A \textbf{576} (1994) 189-204.
\bibitem{Iwasaki} H. Iwasaki \textit{et al.}, Phys. Lett. B \textbf{481} (2000) 7.
\bibitem{Watanabe} S. Watanabe \textit{et al.},  Phys. Rev. C \textbf{89} (2014) 044610.
\bibitem{MAGMG} G. Saxena, M. Kumawat, R. Sharma and Mamta Aggarwal, J. Phys. G: Nucl. Part. Phys. 48 (2021) 125102.
\bibitem{MAGIJNI} Mamta Aggarwal and G. Saxena, Int. J. of Mod. Phys. E \textbf{27}, No. 7 (2018) 1850062.
\bibitem{JUNGHAN} A. Junghans, M. de Jong, H.-G. Clerc, A. Ignatyuk, G.Kudyaev, and K.-H. Schmidt, Nucl. Phys. A 629, 635 (1998).
\bibitem{ripl} T. Belgya et.al., Handbook for calculations of nuclear reaction data, RIPL-2 IAEA-TECDOC-1506 (IAEA, Vienna, 2006). Available online at https://www-nds.iaea.org/RIPL-2/
\bibitem{Tatha} Tathagata Banerjee, S. Nath, and Santanu Pal, Phys. Rev. C 99, 024610 (2019).
\bibitem{MAM} M. Rajasekaran and Mamta Aggarwal, Phys. Rev. {\bf{C 58}} (2004) 2743.
\end{thebibliography}
\end{document}